\documentclass[11pt]{article}
\usepackage{amssymb,amsmath,amsfonts}
\usepackage{graphicx}
\usepackage{graphics}
\usepackage{eepic,epsfig}

1.60cm \makeatletter \@addtoreset{equation}{section}

\makeatother
\begin{document}

\title{Induced fermionic current by a magnetic tube in the cosmic spacetime}
\author{M. S. Maior de Sousa \thanks{E-mail: kaelsousa@gmail.com}\ ,  R. F. Ribeiro \thanks{E-mail: 
rfreire@fisica.ufpb.br} and  E. R. Bezerra de Mello \thanks{E-mail: emello@fisica.ufpb.br} \
\\
Departamento de F\'{\i}sica-CCEN\\
Universidade Federal da Para\'{\i}ba\\
58.059-970, J. Pessoa, PB\\
C. Postal 5.008\\
Brazil}

\maketitle

\begin{abstract}
In this paper, we consider a charged massive fermionic quantum field  in the space-time of an idealized cosmic string, in
the presence of a magnetic field confined in a cylindrical tube of finite radius. Three distinct configurations
for the magnetic field is taken into account: (i) a cylindrical shell of radius $a$, (ii) a magnetic field proportional to $1/r$
and (iii) a constant magnetic field. In these three cases, the axis of the infinitely long
tube of radius $a$ coincides with the cosmic string. Our main objective is to analyze the induced vacuum fermionic
current densities outside the tube. In order to do that, we explicitly construct the wave-functions inside and
outside the tube for each case. Having the complete set of normalized wave-functions, we use the summation method
to develop our analysis. We show that in the region outside the tube, the induced currents are
decomposed into a parts corresponding to a zero-thickness magnetic flux in addition to a core-induced contributions. 
The latter presents specific form depending on the magnetic field configuration considered. We also see that
the only non-vanishing component of fermionic current corresponds to the azimuthal one. The 
zero-thickness contribution depends only on the fractional part of the ration of the magnetic 
flux inside the tube by the quantum one. As to the core-induced contribution, it depends on the 
total magnetic flux inside the tube, and consequently, in general, it is not a periodic function of
the flux. 
\end{abstract}

\bigskip


\bigskip

\section{Introduction}
\label{Int}

The existence of a magnetic flux tube penetrating a type II superconductor,
named {\bf vortex}, was first demonstrated by Abrikosov
\cite{Abrikosov}, by using the Ginzburg-Landau theory of superconductivity.
Some years later, Nielsen and Olesen \cite{nielsen} have shown, 
by using a classical relativistic field theory, composed by Higgs fields
interacting with Abelian one, that presents
spontaneously gauge symmetry broken, contains static cylindrically symmetric
solution carrying a magnetic flux. This configuration corresponds to the vortex solution.
The equations of motion associated for this system form a set of coupled
non-linear differential equation, that, in general, has no closed solutions. The analysis
of the influence of this system on the geometry of the spacetime was analyzed
numerically by Garfinkle  \cite{DG} and Laguna \cite{Laguna} many years ago. In
these papers the authors showed that, the vortex has a inner structure
characterized by a non-vanishing core carrying a magnetic flux, whose
extension is determined by the energy scale where the symmetry is broken. Two
length scales naturally appear, the one related with the extension of the
magnetic flux proportional to the inverse of vector field mass, $m_v$, and
the other associated with the inverse of the scalar field mass, $m_s$. The latter
being the radius that the scalar field reaches its vacuum
value. Moreover the authors also verify that asymptotically the surface
perpendicular to the vortex corresponds to a Minkowski one minus a wedge.
A special vortex solution that satisfy the Bogomolnyi-Prasad-Somerfield (BPS)
limit, presents both masses equal, i.e., $m_v=m_s$. For this case Linet \cite{Linet}
was able to find, in the infinity electric charge and Higgs field
self-coupling limit, exact solution for the metric tensor,
which is determined in terms of the linear energy density of the string.
In this limit, the surface perpendicular to the vortex-line
solution has a conical structure, and the space around this object
corresponds to an idealized cosmic string.

According to the Big Bang theory, the universe has been expanding and
cooling. During its expansion, the spontaneous breaking of
fundamental symmetries leads the universe to a series of phase
transitions. In most interesting model of high-energy physics, the formation
of topological defects such as domain  walls, monopoles, cosmic
string, among others are predicted  to occur \cite{vilenkin}.
These topologically stable structures have a number of interesting
observable consequences, the detection of which would provide an important
link between cosmology and particle physics. The cosmic strings are
among the various type of topological defects, the most studied. Though
the recent observations of the cosmic microwave background radiation have ruled out
 them as the primary source for primordial density perturbations,
 cosmic strings give rise to a number of
 interesting physical effects such as gamma rays bursts \cite{Bere},
 the emission of gravitational waves \cite{Damour} and the generation of
 high-energy cosmic rays \cite{sigl}. String-like defects also appear in a number of condensed matter
systems, including liquid crystals and graphene made structures.

The complete analysis of the behavior of a quantum charged field in the neighborhood of an Abelian
Nielsen and Olesen (NO) string must take into account not only the influence of the geometry of the
spacetime, but also the influence of the magnetic field. Two distinct analysis can be mentioned:
$(i)$ The first one is to consider the string as an idealized linear topological defect,
having a magnetic field running along it. This case can be treated analytically. $(ii)$ The
second approach is to consider the non-zero thickness for the string. Unfortunately
this problem is analytically intractable.
In the idealized model, the conical structure of the spacetime
modifies the zero-point vacuum fluctuation of quantized fields inducing non-vanishing vacuum
expectation values (VEV) for important physical observable.  This analysis
has been developed in many papers for scalar, fermionic and electromagnetic fields (see, for instance,
references given in \cite{Beze06sc}); moreover considering the presence magnetic flux,
additional polarization effects associated with charged quantum fields
take place \cite{Dowk87}-\cite{Site12}. In particular this magnetic flux
induces non-vanishing current densities, $\langle j^\mu\rangle$.
This phenomenon has been investigated for scalar fields in Ref.\cite{Srir01,Site09}. The analysis of
induced fermionic currents in higher-dimensional cosmic string spacetime
in the presence of a magnetic flux have been developed in Ref. \cite{Mello10a}. In all these analysis, the authors
have shown that induced vacuum current densities along the azimuthal direction appear if the ratio of the magnetic
flux by the quantum one has a nonzero fractional part. Moreover, induced fermionic and scalar current
densities in compactified cosmic string spacetimes have been considered in \cite{Saha-Mello} and \cite{Brag}.

Because the analysis of a quantum system in a realistic model for the NO vortex
cannot be exactly analyzed, an intermediate approach is to assume an 
approximated model for this string, that consists
to consider the spacetime produced by the string as being conical everywhere, but having
a non-zero thickness magnetic field surrounding it. In this way,
some improvements were obtained in the calculations of the VEV of some physical observable
when compared with the idealized case. This approach was used in \cite{Spin03,Spin04} to calculate the VEV
of charged scalar and fermionic energy-momentum tensor, respectively.\footnote{Allen et al
\cite{allen} have analyzed the vacuum polarization effect of a massless
scalar field on a cosmic string spacetime, considering generically the effect of
the string’s core through the non-minimal coupling between the scalar field with the geometry.}
More recently the scalar vacuum current induced by a magnetic flux in a cosmic string considering a non-vanishing
core has been developed in \cite{Mello15}. In these calculations the vacuum polarization effects
present two contributions. The first one associated with a zero-thickness magnetic flux. This contribution
is a periodic function of the magnetic flux inside the tube, with the period equal to the quantum flux.
The second contribution, named core-induced contribution, is not a periodic function of the
flux. In fact the core-induced contribution depends on the total magnetic flux inside the core.
Three different configurations of magnetic flux that allow us to obtain exact wave-function
solutions were considered: $(i)$ A magnetic field on a cylindrical shell, $(ii)$ a magnetic field
proportional to $1/r$, and finally $(iii)$ a homogeneous magnetic field inside the tube.

Motivated by these results, we decided in this paper to analyze the induced fermionic current density
in the system defined above, i.e., an idealized cosmic string surrounded by a magnetic field confined
in a tube of finite radius. In order to do that we shall explicitly calculate the wave-functions inside and
outside the tube for each case, and by the summation method we shall develop the calculation
of induced fermionic current.
	
This paper is organized as follow. In section \ref{Geometry} we describe the background geometry of the spacetime and
the configuration of the magnetic fields. We provide the general structure of the 
complete set of normalized positive- and negative-energy fermionic mode functions, which can be 
adapted for each magnetic field configuration. These functions obey the continuity boundary on the surface
of the cylindrical tube of radius $a$. In section \ref{Mode}, by using the mode-summation method, we show
that the VEVs for the charge density, the radial and axial current densities vanish. Only azimuthal current density
is induced. Then we evaluate the renormalized VEV of this current density. We show that the later can be
decomposed in two parts: The first one corresponds to the induced current by a zero-thickness magnetic flux
in the geometry of an idealized cosmic string, and the second one is induced by the non-zero core of the magnetic tube.
The latter is calculated, separately, for the three kinds of magnetic flux considered. The behavior of them
are discussed in various asymptotic regions of the parameters. We also present some plots associated with 
the core-induced azimuthal current exhibiting its behavior as function of the most relevant physical variables.
Our most relevant conclusions are summarized in section \ref{Conclu}.
In the appendix \ref{AppendA} we present some recurrence relations involving Whittaker functions
$M_{\kappa,\mu}(z)$, needed by us to construct the fermionic wave-functions in a simpler form. We would like to 
point out that these relations were not found in the literature. Because these
relations may be useful for anybody else in the future, we decided to explicitly present them.
Throughout the paper we use the units with $G=\hbar=c=1$.

\section{The geometry and the fermionic wave-functions}
\label{Geometry}
The background geometry associated with an idealized cosmic string along the $z$-axis, can be given, by using
cylindrical coordinates, through the line element below:
\begin{equation}
\label{2.1}
ds^2=dt^2 - dr^2 - r^2d\phi^2-dz^2 \ ,
\end{equation}
where the coordinates take values in range $r\geq 0$, $0\leq\phi\leq\phi_0=2\pi/q$, $-\infty\leq t\leq+\infty$ and
$-\infty\leq z\leq+\infty$. The parameter $q$ associated with the planar angle deficit is related to the mass per unit
length of the string, $\mu_0$,  by $q^{-1}=1-4\mu_0$.

The quantum dynamic of a massive charged spinor field in curved space-time and in the
presence of an electromagnetic four-vector potential, $A_\mu$, is described by the Dirac equation,
\begin{equation}
\label{2.2}
i\gamma^{\mu} (\nabla_\mu + ieA_\mu)\psi-m\psi=0, \ \ \nabla_\mu=\partial_\mu + \Gamma_\mu \ ,
\end{equation}
where $\gamma^\mu$ are the Dirac matrices in curved space-time and $\Gamma_\mu$ is the spin
connection. Both matrices are given in terms of the flat space-time Dirac matrices, $\gamma^{(a)}$, by the relations
\begin{equation}
\label{2.3}
\Gamma_\mu=-\frac14\gamma^{(a)}\gamma^{(b)}e^\nu_{(a)}e_{(b)\nu;\mu}, \ \ \gamma^\mu = {e^\mu}_{(a)}\gamma^{(a)} \ .
\end{equation}

In the above expression ${e^\mu}_{(a)}$ represents the basis tetrad satisfying the relation
${e^\mu}_{(a)}{e^\nu}_{(b)}\eta^{ab}=g^{\mu\nu}$, with $\eta^{ab}$ being the Minkowski space-time metric tensor.

The system that we want to analyse takes into account three different configurations of
magnetic fields. They are: $(i)$ A magnetic field on a cylindrical shell of radius $a$,
$(ii)$ a magnetic field proportional to $1/r$, and finally $(iii)$ a homogeneous magnetic
field inside the tube. In these three cases, the axis of the infinitely long
tube of radius $a$ coincides with the cosmic string.
In order to reproduce these three configurations of magnetic field and proceed with our calculations, we consider
the potential vector given by
\begin{equation}
\label{2.01}
A_{\mu}=(0,0,A_\phi(r),0) \ ,
\end{equation}
with
\begin{equation}
\label{2.012}
A_\phi(r)=-\frac{q\Phi}{2\pi}a(r) \ .
\end{equation}
For the first model,
\begin{equation}
\label{2.013}
a(r)=\Theta(a-r) \ .
\end{equation}
For the two second models, we can represent the radial function $a(r)$ by:
\begin{equation}
\label{2.014}
a(r)=f(r)\Theta (a-r)+\Theta (r-a) \ ,
\end{equation}
with
\begin{eqnarray}
\label{2.015}
f(r)=\left\{\begin{array}{cc}
r/a,&\mbox{for the second model} \  ,  \\
r^2/a^2,&\mbox{for the third model} \  .
\end{array}
\right.
\end{eqnarray}
In the expressions above, $\Theta(z)$ represents the Heaviside function, and $\Phi$ the total magnetic flux.

To find the complete set of mode of fermionic wave-functions for the problem under consideration,
we shall use the standard representation of the flat space Dirac matrices:
\begin{equation}
\label{2.4}
\gamma^{(0)}=\left( \begin{array}{cc}
1& 0 \\
0& -1 \end{array} \right), \ \ \gamma^{(a)}=\left( \begin{array}{cc}
0 & \sigma^a \\
-\sigma^a & 0 \end{array} \right), \ a=1, \ 2, \ 3,
\end{equation}
with $\sigma^1, \ \sigma^2 \ \mbox{and} \ \sigma^3$ being the Pauli matrices. We
take the tetrad basis as follow:
\begin{equation}
\label{2.5}
{e^\mu}_{(a)}=\left( \begin{array}{cccc}
1 & 0 & 0 & 0 \\
0 & \cos(q\phi) & -\sin(q\phi)/r & 0 \\
0 & \sin(q\phi) & \cos(q\phi)/r & 0 \\
0 & 0 & 0 & 1 \end{array} \right),
\end{equation}
where the index $(a)$ identifies the rows of the matrix. With this choice, the gamma matrices take the form
\begin{equation}
\label{2.6}
\gamma^0=\gamma^{(0)}=\left( \begin{array}{cc}
1 & 0 \\
0 & -1 \end{array} \right), \ \ \gamma^{l}=\left( \begin{array}{cc}
0 & \sigma^l \\
-\sigma^l & 0 \end{array} \right),
\end{equation}
where we have introduced the $2 \times 2$ matrices for $l=(r, \ \phi, \ z)$:
\begin{equation}
\label{2.7}
\sigma^{r}=\left( \begin{array}{cc}
0 & e^{-iq\phi} \\
e^{iq\phi} & 0 \end{array} \right), \ \sigma^{\phi}=-\frac{i}{r}\left( \begin{array}{cc}
0 & e^{-iq\phi} \\
-e^{iq\phi} & 0 \end{array} \right), \ \sigma^{z}=\left( \begin{array}{cc}
1 & 0 \\
0 & -1 \end{array} \right).
\end{equation}
For this case the spin connection and combination appearing in the Dirac equation we find
\begin{equation}
\label{2.8}
\Gamma_\mu=\frac{1-q}{2}\gamma^{(1)}\gamma^{(2)}\delta^\phi_\mu, \ \ \ \gamma^\mu \Gamma_\mu=\frac{1-q}{2r}\gamma^r.
\end{equation}
Then the Dirac equation take the form
\begin{equation}
\label{2.9}
\left(\gamma^\mu(\partial_\mu + ieA_\mu)+\frac{1-q}{2r}\gamma^r + im\right)\psi=0 \  .
\end{equation}
	
For positive-energy solutions, assuming the time-dependence of the eigenfunctions in the form
$e^{-iEt}$ and decomposing the spinor $\psi$ into the upper $(\psi_{+})$ and lower $(\psi_{-})$
components, we find the following equations
\begin{equation}
\label{2.10}
(E-m)\psi_{+} + i\left[\sigma^l \left(\partial_l +ieA_l\right)+\frac{1-q}{2r}\sigma^r\right]\psi_{-}=0 \  ,
\end{equation}
\begin{equation}
\label{2.11}
(E+m)\psi_{-} + i\left[\sigma^l \left(\partial_l +ieA_l\right)+\frac{1-q}{2r}\sigma^r\right]\psi_{+}=0  \  .
\end{equation}
Substituting the function $\psi_{-}$ from the second equation into the first one, we
obtain the second order differential equation for the spinor $\psi_{+}$:
\begin{eqnarray}
\label{2.12}
\left[r^2\partial^2_r + r \partial_r + \left(\partial_\phi + ieA_\phi -
i\frac{1-q}{2}\sigma^z\right)^2+ \right.\nonumber\\ \left. r^2(\partial^2_z + E^2 - m^2)-
\frac{e}{r}\sigma^z \partial_r A_\phi\right]\psi_{+}&=&0  \   .
\end{eqnarray}
The same equation is obtained for $\psi_{-}$. So, we may say that the general solutions
to $\psi_+$ and $\psi_-$ can be express in terms of the ansatz below,
compatible with the cylindrical symmetry of the physical system,
\begin{equation}
\label{2.13}
\psi_{+}=e^{-ip.x}\left( \begin{array}{cc}
R_{1}(r)e^{iqn_{1}\phi} \\
R_{2}(r)e^{iqn_{2}\phi} \end{array} \right) \  ,
\end{equation}
\begin{equation}
\label{1.14}
\psi_{-}=e^{-ip.x}\left( \begin{array}{cc}
R_{3}(r)e^{iqn_{1}\phi} \\
R_{4}(r)e^{iqn_{2}\phi} \end{array} \right)\  ,
\end{equation}
where $p.x\equiv Et-kz$.

Imposing that our solutions are eigenfunctions of the total angular momentum
along the string, $J_z$,
\begin{equation}
\label{2.15}
\hat{J_z}\psi=\left(-i\partial_\phi+i\frac{q}{2}\gamma^{(1)}\gamma^{(2)}\right)\psi=qj\psi \ ,
\end{equation}
where
\begin{equation}
j=n+1/2, \ n=0, \ \pm 1, \ \pm 2, \ ...   \ ,
\end{equation}
we obtain that $n_2=n_1+1$. From now on we shall use the notations $n_1=n$ and $n_2=n+1$.

In order to construct the complete set of the wave-functions we shall consider, separately, \eqref{2.12}
in the regions $r<a$ and $r>a$. To the first region, $(r<a)$, we have three different configurations of magnetic field
already specified by the four-vector potential \eqref{2.01}-\eqref{2.015}. These configurations 
of the magnetic fields have been used in \cite{Bordag93} to analyze the non-relativistic
quantum motion of a spin $1/2$ charged particle with
a gyromagnetic ratio $g\neq 2$ interacting with magnetic field and considering the presence of a
magnetic dipole interaction; moreover, the corresponding relativistic analysis has bee developed in \cite{Bordag94}.
In the latter, only two configurations of magnetic field have been considered, the homogeneous field and
the cylindrical shell with a delta-Dirac distribution.

For the outside region, $r>a$, there is no magnetic field and the vector potential given by,
\begin{equation}
\label{A-B}
A_\phi=-\frac{q\Phi}{2\pi} \  ,
\end{equation}
being $\Phi$ the magnetic flux. So, the equation (\ref{2.12}) can written as following:
\begin{eqnarray}
\label{2.18}
\left[r^2\partial^2_r + r \partial_r + \left(\partial_\phi + ieA_\phi - i\frac{1-q}{2}
\sigma^z\right)^2\ + \right.\nonumber\\ \left. r^2(\partial^2_z + E^2 - m^2)\right]\psi_{+}=0 \  .
\end{eqnarray}
Substituting (\ref{A-B}) into (\ref{2.18}) we find that the external positive-energy solution of the Dirac equation
is given in terms of Bessel, $J_\mu(z)$, and Numann, $Y_\mu(z)$, functions. A similar
result is verified for the lower components of the Dirac spinor. Consequently we can write:
\begin{equation}
\label{2.19}
\psi^{(+)}=e^{-ip.x}e^{iqn\phi}\left( \begin{array}{c}
C_{1}J_{\beta_j}(\lambda r)+D_{1}Y_{\beta_j}(\lambda r) \\
\left[C_{2}J_{\beta_j+\epsilon_j}(\lambda r)+D_{2}Y_{\beta_j+\epsilon_j}(\lambda r)\right]e^{iq\phi} \\
A_{1}J_{\beta_j}(\lambda r)+B_{1}Y_{\beta_j}(\lambda r) \\
\left[A_{2}J_{\beta_j+\epsilon_j}(\lambda r)+B_{2}Y_{\beta_j+\epsilon_j}(\lambda r)\right]e^{iq\phi} \end{array} \right),
\end{equation}
with $n=j-1/2$ being an integer number. We have defined $\epsilon_j=1$ for $j\geq-\alpha$ and
$\epsilon_j=-1$ for $j<-\alpha$, being $\alpha=eA_\phi/q=-\Phi/\Phi_0$, Here $\Phi_0=2\pi/e$ is the quantum flux, and
\begin{equation}
\label{2.20}
\beta_j=q|j+\alpha|-\frac{\epsilon_j}{2}.
\end{equation}
As we can see, we have introduced a set of eight arbitrary constants $C_i$, $D_i$, $A_i$ and
$B_i$ with $i=1, \ 2$ in the general solution above.

The energy is expressed in terms of $\lambda, \ k \ \mbox{and} \ m$ by the relation
\begin{equation}
\label{2.21}
E=\sqrt{\lambda^2 + k^2 + m^2}.
\end{equation}
We can find a relation between the constants of the the upper and lower
solutions in (\ref{2.19}) by the use of (\ref{2.10}) and (\ref{2.11}). The relations are given by
\begin{equation}
\label{2.22}
A_1=\frac{kC_1-i\epsilon_j \lambda C_2}{E+m}, \ \ A_2=-\frac{kC_2-i\epsilon_j \lambda C_1}{E+m}
\end{equation}
\begin{equation}
\label{2.23}
B_1=\frac{kD_1-i\epsilon_j \lambda D_2}{E+m}, \ \ B_2=-\frac{kD_2-i\epsilon_j \lambda D_1}{E+m}.
\end{equation}
In addition, for the further specification of the eigenfunctions, we
can impose extra conditions relating the above constants.
As such a condition, following \cite{Nail}, we will
require the following relations between the upper and
lower components:
\begin{equation}
\label{2.24}
R_3(r)=\rho_s R_1(r), \ \ R_4(r)=-\frac{R_2(r)}{\rho_s},
\end{equation}
with
\begin{equation}
\label{rho}
\rho_s=\frac{E+s\sqrt{\lambda^2 + m^2}}{k} \ , \  s= \pm 1  \ .
\end{equation}
Doing this we obtain the following relations:
\begin{eqnarray}
\label{rho1}
A_1=\rho_sC_1 \ , \ A_2=-C_2/\rho_s \  , \\
B_1=\rho_sD_1 \ ,  \  B_2=-D_2/\rho_s \  .
\end{eqnarray}
Hence, the positive frequency exterior solutions to the Dirac
equation, specified by the set of quantum numbers $\sigma=(\lambda, \ j, \ k, \ s)$,  has the form
\begin{equation}
\label{out}
\psi^{(+)}_{\sigma(out)}(x)=e^{-ip.x}e^{iqn\phi}\left( \begin{array}{c}
C_{1}J_{\beta_j}(\lambda r)+D_{1}Y_{\beta_j}(\lambda r) \\
i\epsilon_j \rho_s b^{(+)}_{s}\left[C_{1}J_{\beta_j+\epsilon_j}(\lambda r)+
D_{1}Y_{\beta_j+\epsilon_j}(\lambda r)\right]  e^{iq\phi} \\
\rho_s\left[C_{1}J_{\beta_j}(\lambda r)+D_{1}Y_{\beta_j}(\lambda r)\right] \\
-i\epsilon_j b^{(+)}_{s}\left[C_{1}J_{\beta_j+\epsilon_j}(\lambda r)+
D_{1}Y_{\beta_j+\epsilon_j}(\lambda r)\right] e^{iq\phi} \end{array} \right),
\end{equation}
where we have introduced
\begin{equation}
\label{2.26}
b^{(\pm)}_s=\frac{\pm m+s\sqrt{\lambda^2+m^2}}{\lambda} \  .
\end{equation}

For the region inside, $r<a$, we have three different configurations of magnetic field,
as we have already mentioned. In this way we have three different solutions for
\eqref{2.12}. Let us represent each radial function by $R_l^{(i)}$, where $i=1, \ 2, \ 3$,
is the index associated with the model and $l=1, \ 2, \ 3, \ 4$  the index specifying the
the spinor components. By using the general expression for the azimuthal
component of the four-vector potential \eqref{2.01}, we can shown that the relations \eqref{2.24}
and \eqref{rho} still hold. So, using the relations \eqref{2.10} and \eqref{2.11} between
the upper and lower components of the spinor fields, we can write the inner positive-energy
spinor field in the general form below:
\begin{equation}
\label{in}
\psi^{(+)}_{i(in)}(x)=C^{(i)} e^{-ip.x}e^{iqn\phi}\left( \begin{array}{c}
R^{(i)}_1(\lambda, r)\\
i\rho_s b^{(+)}_{s} R^{(i)}_2(\lambda, r) e^{iq\phi} \\
\rho_s R^{(i)}_1(\lambda, r) \\
-i b^{(+)}_{s}R^{(i)}_2(\lambda, r) e^{iq\phi} \end{array} \right) \   .
\end{equation}

The coefficients $C_1$ and $D_1$ in \eqref{out} and $C^{(i)}$ in \eqref{in} are determined from
the continuity condition of the fermionic wave function at $r=a$. After some intermediate
steps we can write,
\begin{eqnarray}
\label{Rela1}
C_1&=&-\frac\pi2(\lambda a)C^{(i)}R^{(i)}_1(\lambda, a){\tilde{Y}}_{\beta_j}(\lambda a) \ , \\
\label{Rela2}
D_1&=&\frac\pi2(\lambda a)C^{(i)}R^{(i)}_1(\lambda, a){\tilde{J}}_{\beta_j}(\lambda a) \  ,
\end{eqnarray}
where
\begin{eqnarray}
\label{Z.Bessel}
{\tilde{Z}}_{\beta_j}(z)=\epsilon_jZ_{\beta_j+\epsilon_j}(z)-{\cal{V}}_j^{(i)}(\lambda, a)
 Z_{\beta_j} (z)  \ , \  {\rm with} \
{\cal{V}}^{(i)}_j(\lambda, a)=\frac{R_2^{(i)}(\lambda, a)}{R^{(i)}_1(\lambda, a)} \ .
\end{eqnarray}
In \eqref{Z.Bessel} $Z_\mu$ represents the Bessel functions $J_\mu$ or $Y_\mu$.
With this notation all the informations about the inner structure of the magnetic field is contained in
the coefficient ${\cal{V}}_j^{(i)}$.

Finally the constant $C^{(i)}$ can be obtained form the normalization condition,
\begin{equation}
\label{Ren}
\int{d^3x\sqrt{g^{(3)}}\left(\psi^{(+)}_\sigma\right)^\dagger\psi^{(+)}_{\sigma'}}=
\delta_{\sigma,\sigma'} \ ,
\end{equation}
where delta symbol on the right-hand side is understood as the Dirac delta function for
continuous quantum numbers $\lambda \ \mbox{and} \ k$, and the Kronecker delta for discrete ones
 $n \ \mbox{and} \ s$, and $g^{(3)}$ is the determinant of the spatial metric tensor.
 The integral over the radial coordinate should be done in the interval $[0, \ \infty)$.
Due to cylindrical symmetry of this system we can write the spinor field in a general form
as shown bellow:
 \begin{equation}
\label{General}
\psi^{(+)}_{\sigma}(x)=e^{-ip.x}e^{iqn\phi}\left( \begin{array}{c}
F_1(\lambda, r)\\
i\rho_s b^{(+)}_{s} F_2(\lambda, r) e^{iq\phi} \\
\rho_s F_1(\lambda, r) \\
-i b^{(+)}_{s}F_2(\lambda, r) e^{iq\phi} \end{array} \right) \   .
\end{equation}
Consequently from \eqref{Ren} we obtain,
\begin{eqnarray}
\label{IntF}
(1+\rho_s^2)\left[\int_0^\infty dr \ r F_1^*(\lambda, r)F_1(\lambda', r)
+(b_s^{(+)})^2\int_0^\infty dr \ r F_2^*(\lambda, r)F_2(\lambda', r)\right]=
\frac q{(2\pi)^2}\delta(\lambda-\lambda') \  . \nonumber \\
\end{eqnarray}
The integral over the interior region is finite, consequently the dominant contribution
in \eqref{IntF} for $\lambda'=\lambda$ comes from the integration in the exterior region.
By using the standard integrals involving the cylindrical Bessel
functions, we find
\begin{equation}
\label{Rela3}
(2\pi)^2[|C_1|^2+|D_1|^2]=\frac{q\lambda}{(1+\rho^2_s)(1+({b^{(+)}_s})^2)}  \  .
\end{equation}
Substituting  \eqref{Rela1} and \eqref{Rela2} into the above equation we find:
\begin{eqnarray}
\label{Rela4}
C^{(i)}R^{(i)}_1(\lambda, a)=\Xi(\lambda,a) \  ,
\end{eqnarray}
with
\begin{eqnarray}
\Xi(\lambda,a)=\frac1{a\pi^2}\left[\frac q\lambda\frac1{(1+\rho^2_s)}
\frac1{(1+({b^{(+)}_s})^2)}\right]^{1/2} \frac1{\sqrt{(\tilde{Y}_{\beta_j}(\lambda a))^2
+(\tilde{J}_{\beta_j}(\lambda a))^2}} \  .
\end{eqnarray}
This relation determines the normalization constant for the interior wave function. With this
normalization we can construct the spinor function for the external region, $r>a$:
\begin{equation}
\label{psi-out}
\psi^{(+)}_{\sigma(out)}(x)=N^{(+)}e^{-ip.x}e^{iqn\phi}\left( \begin{array}{c}
g_{\beta_j}(\lambda a, \lambda r) \ ,  \\
i\epsilon_j \rho_s b^{(+)}_s g_{\beta_j+\epsilon_j}(\lambda a, \lambda r) e^{iq\phi} \\
\rho_s g_{\beta_j}(\lambda a, \lambda r) \\
-i\epsilon_j b^{(+)}_s g_{\beta_j+\epsilon_j}(\lambda a, \lambda r) e^{iq\phi} \end{array} \right),
\end{equation}
where we have introduced the notations,
\begin{equation}
\label{2.36}
N^{(+)}=\frac{1}{2\pi}\left[\frac{q\lambda}{(1+\rho^2_s)\left(1+(b^{(+)}_s)^2\right)}\right]^{1/2}  \  ,
\end{equation}
\begin{equation}
\label{g-beta}
g_{\beta_j}(\lambda a, \lambda r)=\frac{\tilde{Y}_{\beta_j}(\lambda a)J_{\beta_j}(\lambda r)
-\tilde{J}_{\beta_j}(\lambda a)Y_{\beta_j}(\lambda r)}{\sqrt{(\tilde{Y}_{\beta_j}(\lambda a))^2
+(\tilde{J}_{\beta_j}(\lambda a))^2}}
\end{equation}
and
\begin{equation}
\label{g-beta1}
g_{\beta_j+\epsilon_j}(\lambda a, \lambda r)=\frac{\tilde{Y}_{\beta_j}(\lambda a)
J_{\beta_j+\epsilon_j}(\lambda r)-\tilde{J}_{\beta_j}(\lambda a)Y_{\beta_j+\epsilon_j}
(\lambda r)}{\sqrt{(\tilde{Y}_{\beta_j}(\lambda a))^2+(\tilde{J}_{\beta_j}(\lambda a))^2}}  \   .
\end{equation}

The out-side negative-energy fermionic wave-function can be obtained in a similar way, or
by the charge conjugation operator, $\psi^{(-)}_\sigma=i\gamma^{(2)}\hat{K}\psi^{(+)}_\sigma$,
where $\hat{K}$ is the complex conjugation operator \cite{B-D}. The corresponding
result is given by the expression:
\begin{equation}
\label{psi-out1}
\psi^{(-)}_{\sigma(out)}(x)=N^{(-)}e^{ip.x}e^{-iq(n+1)\phi}\left( \begin{array}{c}
g_{\beta_j+\epsilon_j}(\lambda a, \lambda r) \\
i\epsilon_j \rho_s b^{(-)}_s g_{\beta_j}(\lambda a, \lambda r) e^{iq\phi} \\
\rho_s g_{\beta_j+\epsilon_j}(\lambda a, \lambda r) \\
-i\epsilon_j b^{(-)}_s g_{\beta_j}(\lambda a, \lambda r) e^{iq\phi} \end{array} \right),
\end{equation}
where the phase $i\epsilon_j$ was absorbed into the normalization constant and
have used $b^{(-)}_s=1/b^{(+)}_s$. The
normalization constant becomes,
\begin{equation}
\label{norm}
N^{(-)}=(N^{(+)})^\dagger b^{(+)}_s=\frac{1}{2\pi}\left[{\frac{q\lambda}{(1+\rho^2_s)
\left(1+(b^{(-)}_s)^2\right)}}\right]^{1/2}  \  .
\end{equation}

Having the negative-energy wave-function for the region outside the magnetic flux, i.e.,
for $r>a$, we can use it for the investigation
of the vacuum fermionic current densities.

\section{Fermionic current}
\label{Mode}
Here we shall evaluate the vacuum expectation value (VEV) of the fermionic current density
operator, $j^{\mu}=e\bar{\psi}\gamma^{\mu}\psi$,
by using the mode sum formula,
\begin{equation}
\label{current}
\left\langle j^\mu(x) \right\rangle=e\sum_{\sigma}{\bar{\psi}^{(-)}_\sigma (x)\gamma^\mu {\psi}^{(-)}_\sigma (x)}  \    ,
\end{equation}
where we are using the compact notation defined by
\begin{equation}
\label{sum}
\sum_{\sigma}=\int^{\infty}_{0} d\lambda \int^{+\infty}_{-\infty} dk \sum_{j=\pm 1/2, ...} \sum_{s=\pm 1}  \  .
\end{equation}

In the following subsections  we shall calculate separately all the components
of the VEV of the fermionic current densities.

\subsection{Charge and radial current densities}
\label{ssec1}

Let us start the calculation of the charge density,
\begin{equation}
\label{charge}
\rho(x)=\left\langle j^0(x) \right\rangle=e\sum_{\sigma}{(\psi^{(-)}_\sigma (x))^\dagger {\psi}^{(-)}_\sigma (x)} \  .
\end{equation}
Substituting \eqref{psi-out1} and \eqref{norm} into \eqref{charge} we obtain
\begin{equation}
\label{charge0}
\rho(x)=\frac{eq}{(2\pi)^2}\int_{-\infty}^\infty dk \int_0^\infty d\lambda \sum_{j=\pm 1/2, ...}
\left((\\g_{\beta_j}(\lambda a, \lambda r))^2+(g_{\beta_j+\epsilon_j}(\lambda a, \lambda r))^2\right) \ .
\end{equation}
Moreover, substituting \eqref{g-beta} and \eqref{g-beta1} into the above expression, and after some
intermediate steps, we obtain,
\begin{equation}
\label{charge1}
\rho(r)=\rho_s(r)+\rho_c(r) \  ,
\end{equation}
where
\begin{equation}
\label{rho-s}
\rho_s(r)=\frac{eq}{(2\pi)^2}\int^{\infty}_{-\infty}dk\sum_{j=\pm 1/2, ...}\int^{\infty}_{0}d\lambda
\lambda\left(J^2_{\beta_j}(\lambda r)+J^2_{\beta_j+\epsilon_j}(\lambda r)\right)
\end{equation}
represents the charge density in a cosmic string spacetime carrying a zero-thickness
magnetic flux along its axis, and
\begin{equation}
\label{rho-c}
\rho_c(r)=-\frac{eq}{2(2\pi)^2}\int^{\infty}_{-\infty}{dk}\sum_{j}\int^{\infty}_{0}
{d\lambda \lambda \tilde{J}_{\beta_j}(\lambda a) \sum^{2}_{l=1}{\frac{(H^{(l)}_{\beta_j}
(\lambda r))^2+(H^{(l)}_{\beta_j+\epsilon_j}(\lambda r))^2}{\tilde{H}^{(l)}_{\beta_j}(\lambda a)}}}  \  ,
\end{equation}
is the charge density induced by the magnetic flux tube around the string. In \eqref{rho-c}
$H^{(l)}_{\nu}(x)$ with $l=1, \ 2$ are the Hankel functions. 

In \cite{Mello13} the charge density, $\rho_s$, was analyzed. There it was
shown that the integrations over $\lambda$ and $k$ in \eqref{rho-s} are divergent. 
In order to obtain a finite and well defined
result we introduced a cutoff function. With this cutoff function the integrals could be evaluated.
By subtracting the Minkowskian part, which corresponds to subtract $\alpha_0=0$ and $\ q=1$ contribution in the
result, the cutoff function can be removed and
a vanishing result was found for the renormalized charge density.

As to $\rho_c$, in principle it is finite and does not require any regularization procedure. To evaluate
this contribution we proceedings as follows: in the complex $\lambda$ plane we rotate the integration
contour by the angle $\pi/2$ for $l=1$  and by the angle $-\pi/2$ for $l=2$. Moreover, it shown in
Appendix \ref{AppendA} that the coefficient ${\cal{V}}^{(i)}_j(\lambda, a)$ in \eqref{Z.Bessel}
satisfy the relation below:\footnote{ In fact the relation \eqref{Rel-V} is satisfied
for all radial functions associated with the three configurations of magnetic field.}
\begin{equation}
\label{Rel-V}
{\cal{V}}^{(i)}_j(\pm i\lambda, a)=\pm i{\rm Im}\{{\cal{V}}^{(i)}_j( i\lambda, a)\}  \  .
\end{equation}
 So, by using \eqref{Rel-V} and the well known relations
involving the Bessel and Hankel functions of imaginary argument with the modified Bessel
functions \cite{Abramo}, we can see that the integrand of the first contribution $(l=1)$ cancels
the other from the second contribution $(l=2)$, providing a vanishing result for \eqref{rho-c}.
So, we conclude that there is no induced charge density for this system.

To analyze the VEV of the radial and axial current densities, we use the sum mode below,
\begin{equation}
\label{radial}
\left\langle j^r(x) \right\rangle=e\sum_{\sigma}{(\psi^{(-)}_\sigma (x))^\dagger\gamma^0
\gamma^r {\psi}^{(-)}_\sigma (x)}
\end{equation}
and
\begin{equation}
\label{Axial}
\left\langle j^z(x) \right\rangle=e\sum_{\sigma}{(\psi^{(-)}_\sigma (x))^\dagger
\gamma^0 \gamma^z {\psi}^{(-)}_\sigma (x)}  \  .
\end{equation}
Substituting \eqref{psi-out1} and \eqref{norm} into the above expressions we observe that:
For the radial current density there is a direct cancellations between all terms
involved. As to the axial current the result obtained is an odd function
of the variable $k$, consequently integrating over this variable a vanishing result
is promptly obtained. This latter result is in agreement with the boost symmetry
along the $z$-direction of the system.  So, we  conclude that there are no induced
radial or axial current densities in this system.

\subsection{Azimuthal current density}

The VEV of the azimuthal current density is given by,
\begin{eqnarray}
\label{Azimuthal}
\langle j^\phi (x) \rangle=e\sum_{\sigma}{(\psi^{(-)}_\sigma (x))^\dagger
\gamma^0 \gamma^\phi {\psi}^{(-)}_\sigma (x)}  \  .
\end{eqnarray}
By using \eqref{psi-out1} and \eqref{norm} and the explicit form of the Dirac
matrices given by (\ref{2.6}) and (\ref{2.7}) the following expression below is obtained:
\begin{eqnarray}
\label{Azimuthal1}
\langle j^\phi \rangle=\frac{eq}{2 \pi^2 r}\sum_{\sigma}\frac{\epsilon_j b^{(-)}_s
(\rho^2_s - 1) \lambda}{(1+(b^{(-)}_s)^2)(1+\rho^2_s)}g_{\beta_j}(\lambda a, \lambda r)
g_{\beta_j+\epsilon_j}(\lambda r, \lambda r)  \   .
\end{eqnarray}
We can easily see that
\begin{eqnarray}
\frac{b^{(-)}_s (\rho^2_s - 1)}{(1+(b^{(-)}_s)^2)(1+\rho^2_s)}=
\frac{\lambda}{2\sqrt{\lambda^2 + k^2 +m^2}} \  .
\end{eqnarray}
So, substituting this expression into \eqref{Azimuthal1}, we can see that the summation over $s$
provides a factor $2$. So, it remains only the expression below,
\begin{eqnarray}
\label{Azimuthal2}
\langle j^\phi \rangle=\frac{eq}{2\pi^2 r}\int^{\infty}_{-\infty} dk \int^{\infty}_{0}
\frac{\lambda^2d\lambda}{\sqrt{\lambda^2 + k^2 + m^2}}\sum_{j}\epsilon_j
g_{\beta_j}(\lambda a, \lambda r)g_{\beta_j+\epsilon_j}(\lambda a, \lambda r) \   .
\end{eqnarray}
Developing the product of $g_{\beta_j}(\lambda a, \lambda r)g_{\beta_j+\epsilon_j}(\lambda a, \lambda r)$
in a convenient form, i.e., separating the contributions that does not depend on the 
inner structure of the magnetic field from the other that does, we can written
the above result as the sum of two terms as shown below:
\begin{eqnarray}
\label{Azimuthal3}
\langle j^\phi (x) \rangle=\langle j^\phi (x) \rangle_{s} + \langle j^\phi (x) \rangle_{c} \  .
\end{eqnarray}
The first term, $\langle j^\phi (x) \rangle_{s}$, corresponds to the azimuthal current density in
the geometry of a straight cosmic having a magnetic flux running along its core,
and the second, $\langle j^\phi (x) \rangle_{c}$, is induced by the magnetic tube of radius $a$.

At this point we would like to analyze separately both contributions.

\subsubsection{Azimuthal current induced by a zero-thickness magnetic flux}

The azimuthal current induced by a magnetic flux running along the 
idealized cosmic string is given by,
\begin{eqnarray}
\label{Azimu}
\langle j^\phi (x) \rangle_{s}=\frac{eq}{2\pi^2 r}
\int^{\infty}_{-\infty} dk \int^{\infty}_{0}\frac{d\lambda \lambda^2}
{\sqrt{\lambda^2 + k^2 + m^2}}\sum_{j}\epsilon_j J_{\beta_j}(\lambda r)J_{\beta_j+\epsilon_j}(\lambda r) \  .
\end{eqnarray}

The explicit calculation of this contribution was given in \cite{Mello13}. Here we briefly review 
its more important results. Using the identity bellow, 
\begin{equation}
\frac{1}{\sqrt{m^{2}+k^{2}+\lambda ^{2}}}=\frac{2}{\sqrt{\pi }}%
\int_{0}^{\infty }dt\ e^{-(m^{2}+k^{2}+\lambda ^{2})t^{2}} \  ,  \label{ident1}
\end{equation}
into \eqref{Azimu}, it is possible to integrate over the variable $\lambda$
by using the results form \cite{Grad}:
\begin{equation}
\int_{0}^{\infty }d\lambda \lambda ^{2}e^{-\lambda ^{2}t^{2}}J_{\beta
_{j}}(\lambda r)J_{\beta _{j}+\epsilon _{j}}(\lambda r)=\frac{%
e^{-r^{2}/(2t^{2})}}{4t^{4}}r\epsilon _{j}\left[ I_{\beta
_{j}}(r^{2}/(2t^{2}))-I_{\beta _{j}+\epsilon _{j}}(r^{2}/(2t^{2}))\right] \ .
\label{Int1}
\end{equation}
Introducing a new variable $y=r^{2}/(2t^{2})$, we get
\begin{equation}
\langle j^{\phi }\rangle _{s}=\frac{eq}{2\pi ^{2}r^{4}}\int_{0}^{\infty }\
dy\ y\ e^{-y-m^{2}r^{2}/(2y)}\ [\mathcal{I}(q,\alpha _{0},y)-\mathcal{I}%
(q,-\alpha _{0},y)]\ ,  \label{j-cs1}
\end{equation}
where $\mathcal{I}(q,\alpha _{0},y)$ is defined by
\begin{equation}
\mathcal{I}(q,\alpha _{0},z)=\sum_{j}I_{\beta _{j}}(z)=\sum_{n=0}^{\infty }%
\left[ I_{q(n+\alpha _{0}+1/2)-1/2}(z)+I_{q(n-\alpha _{0}+1/2)+1/2}(z)\right]
,  \label{seriesI0}
\end{equation}%
and
\begin{equation}
\sum_{j}I_{\beta _{j}+\epsilon _{j}}(z)=\mathcal{I}(q,-\alpha _{0},z) \ .
\label{seriesI2}
\end{equation}
In the above development, we have used used the notation
\begin{eqnarray}
\label{alpha}
\alpha=eA_\phi/q =-\Phi/\Phi_0= n_0+\alpha_0 \  ,
\end{eqnarray}
with $n_0$ being an integer number. So we conclude that \eqref{j-cs1}
is an odd function of $\alpha_0$.

In \cite{Mello10} we have presented an integral representation for $\mathcal{I}(q,\alpha _{0},y)$:
\begin{eqnarray}
&&\mathcal{I}(q,\alpha _{0},z)=\frac{e^{z}}{q}-\frac{1}{\pi }%
\int_{0}^{\infty }dy\frac{e^{-z\cosh y}f(q,\alpha _{0},y)}{\cosh (qy)-\cos
(q\pi )}  \notag \\
&&\qquad +\frac{2}{q}\sum_{k=1}^{p}(-1)^{k}\cos [2\pi k(\alpha
_{0}-1/2q)]e^{z\cos (2\pi k/q)},  \label{seriesI3}
\end{eqnarray}
with $2p<q<2p+2$ and with the notation%
\begin{eqnarray}
f(q,\alpha _{0},y) &=&\cos \left[ q\pi \left( 1/2-\alpha _{0}\right) \right]
\cosh \left[ \left( q\alpha _{0}+q/2-1/2\right) y\right]  \notag \\
&&-\cos \left[ q\pi \left( 1/2+\alpha _{0}\right) \right] \cosh \left[
\left( q\alpha _{0}-q/2-1/2\right) y\right] \ .  \label{fqualf}
\end{eqnarray}
For $1\leqslant q<2$, the last term on the right-hand side of Eq. (\ref{seriesI3}%
) is absent.

By using the result (\ref{seriesI3}), and after the integration over $y$, the
expression (\ref{j-cs1}) is presented in the form
\begin{eqnarray}
\langle j^{\phi }\rangle _{s} &=&\frac{em^{2}}{\pi ^{2}r^{2}}\ \left[
\sum_{k=1}^{p}\frac{(-1)^{k}}{\sin (\pi k/q)}\sin (2\pi k\alpha
_{0})K_{2}(2mr\sin (\pi k/q))\right.   \notag \\
&&\left. +\frac{q}{\pi }\int_{0}^{\infty }dy\frac{g(q,\alpha
_{0},2y)K_{2}(2mr\cosh y)}{[\cosh (2qy)-\cos (q\pi )]\cosh y}\right] \ .
\label{jazimu}
\end{eqnarray}
In the above expression $K_{\nu }(x)$ is the Macdonald function, and
\begin{eqnarray}
g(q,\alpha _{0},y) &=&\cos \left[ q\pi \left( 1/2+\alpha _{0}\right) \right]
\cosh \left[ q\left( 1/2-\alpha _{0}\right) y\right]   \notag \\
&&-\cos \left[ q\pi \left( 1/2-\alpha _{0}\right) \right] \cosh \left[
q\left( 1/2+\alpha _{0}\right) y\right] .  \label{gxy}
\end{eqnarray}

As we can see $\langle j^{\phi }\rangle _{s}$ depends only on $\alpha_0$, and vanishes
for the case where this parameter is zero

\subsubsection{Core-induced azimuthal current}

As to $\langle j^\phi (x) \rangle_{c}$, we can show that it can be written by,
\begin{eqnarray}
\label{Azimu1}
\langle j^\phi (x) \rangle_{c}&=&-\frac{eq}{(2\pi)^2 r}\int^{\infty}_{-\infty} dk
\int^{\infty}_{0}\frac{d\lambda \lambda^2 }{\sqrt{\lambda^2 + k^2 + m^2}} \nonumber \\
&&\times\sum_{j}\epsilon_j \tilde{J}_{\beta_j}(\lambda a)\sum^{2}_{l=1}\frac{H^{(l)}_{\beta_j}
(\lambda r)H^{(l)}_{\beta_j+\epsilon_j}(\lambda r)}{\tilde{H}^{(l)}_{\beta_j}(\lambda a)}  \   .
\end{eqnarray}
In order to develop this calculation, we rotate the integrals contour in
the complex plane $\lambda$ as follows: by the angle $\pi/2$ for $l=1$
and $-\pi/2$ for $l=2$. By using the property \eqref{Rel-V}, one can see that the integral over the segments
$(0, \ i\sqrt{m^2+k^2})$ and $(0, \ -i\sqrt{m^2+k^2})$ are canceled. In the remaining integral over the
imaginary axis we introduce the modified Bessel functions. Moreover, writing imaginary integral
variable by $\lambda=\pm iz$, the core-induced azimuthal current reads,
\begin{eqnarray}
\label{Azimu2}
\langle j^\phi (x)\rangle_{c}&=&\frac{eq}{\pi^3 r}\int_0^\infty dk \int^{\infty}_{\sqrt{k^2 + m^2}}
\frac{dz z^2}{\sqrt{z^2 - k^2 - m^2}} \nonumber \\
 &&\sum_{j}K_{\beta_j}(z r)K_{\beta_j + \epsilon_j}(z r)F^{(i)}_{j}(z a) \  ,
\end{eqnarray}
where we use the notation
\begin{eqnarray}
\label{F-ratio}
F^{(i)}_{j}(y)= \frac{I_{\beta_j+\epsilon_j}(y)-
\mbox{Im}[{\cal{V}}_j^{(i)}(iy/a, a)]I_{\beta_j}(y)}{K_{\beta_j+\epsilon_j}(y)+
\mbox{Im}[{\cal{V}}_j^{(i)}(iy/a, a)]K_{\beta_j}(y)} \  .
\end{eqnarray}
After a convenient coordinate transformations we write \eqref{Azimu2} as follow:
\begin{eqnarray}
\label{Azimu3}
\langle j^\phi (x)\rangle_{c}=\frac{eq}{\pi^2 r^4}\int^{\infty}_{mr}z^2 dz \sum_{j} K_{\beta_j}(z)
K_{\beta_j + \epsilon_j}(z)F^{(i)}_{j}(z (a/r)) \  .
\end{eqnarray}

Before to start explicit numerical analysis related to the core-induced 
azimuthal current, let us now evaluate the behavior of it at large distance
from the core. First we consider the massive fields and in the limit $mr>>1$.  
In order to develop this analysis, we assume that the product 
$K_{\beta_j}(z)K_{\beta_j + \epsilon_j}(z)$ can be expressed in terms of 
their corresponding asymptotic forms. So we have the induced current density given below
\begin{equation}
\label{massive2}
\langle j^\phi\rangle_c \approx \frac{eq}{2\pi r^4}\int^{\infty}_{mr} 
dz \ z e^{-2z}\sum_{j}F^{(i)}_{j}(z (a/r)) \  .
\end{equation}
The dominant contribution is given from the region near the lower limit of integration. 
Then the leading order contribution is,
\begin{equation}
\label{massive4}
\langle j^\phi(r)\rangle_c \approx \frac{eqm^4}{4\pi (mr)^3}e^{-2mr}\sum_{j}F^{(i)}_{j}(m a) \  .
\end{equation}
So, we can see that for massive fields and at large distances from the core, the 
core-induced azimuthal current decays with $e^{-2mr}/(mr)^3$. 
Comparing this behavior with the corresponding
one for the zero-thickness azimuthal current, $\left\langle j_{\phi }\right\rangle _{c}$, 
given by Eq. \eqref{jazimu}, we observe in \cite{Mello13} that for $q>2$ the latter 
decays with $e^{-2mr\sin(\pi/q)}/(mr)^{5/2}$, consequently dominates the 
total azimuthal current. For $q\leqslant 2$, the
contributions of $\left\langle j_{\phi }\right\rangle _{s}$ and 
$\left\langle j_{\phi }\right\rangle _{c}$ to the total azimuthal current, at large
distances, are of the same order.

Our next analysis will be developed by using explicitly the radial functions in the region 
inside the tube. In appendix \ref{AppendA} we
provide exact solutions for $R_1(r)$ and $R_2(r)$ for the three different configurations
of magnetic field. They are:
\begin{enumerate}
\item For the cylindrical shell:
\begin{eqnarray}
\label{30.01}
R^{(1)}_1(r)&=& J_{\nu_j}(\lambda r)\nonumber\\
R^{(1)}_2(r)&=&\hat{\epsilon}_j J_{\nu_j+\hat{\epsilon}_j}(\lambda r) \ ,
\end{eqnarray}
 where $\nu_j=q|j|-\frac{\tilde{\epsilon}_j}{2}$, with $\tilde{\epsilon}_j=1$ for $j>0$ and
$\tilde{\epsilon}_j=-1$ for $j<0$.

\item For the magnetic field proportional to $1/r$:
\begin{eqnarray}
\label{30.02}
R^{(2)}_1(r)&=&\frac{M_{\kappa,\nu_j}(\xi r)}{\sqrt{r}}\nonumber\\
R^{(2)}_2(r)&=&C^{(2)}_j \frac{M_{\kappa,\nu_j+\hat{\epsilon}_j}(\xi r)}{\sqrt{r}} \  ,
\end{eqnarray}
where
\begin{eqnarray}
\label{2.300}
\xi=\frac{2}{a}\sqrt{q^2\alpha^2 -\lambda^2 a^2} \  , \ \kappa=-\frac{2q^2 j \alpha}{\xi a}
\end{eqnarray}
and
\begin{equation}
\label{2.30.0}
C^{(2)}_j=\left\{\begin{array}{cc}
\frac{\lambda}{\xi}\frac{1}{(2q|j|+1)} ,  \  \ j>0  \ .
\\ -\frac{\xi}{\lambda}(2q|j|+1), \ j<0 \ . \end{array} \right.
\end{equation}
\item For the homogeneous magnetic field:
\begin{eqnarray}
\label{30.03}
R^{(3)}_1(r)&=&\frac{M_{\kappa-\frac{1}{4},\frac{\nu_j}{2}}(\tau r^2)}r \nonumber\\
R^{(3)}_2(r)&=&C^{(3)}_j \frac{M_{\kappa+\frac{1}{4},\frac{\nu_j+\hat{\epsilon}_j}{2}}(\tau r^2)}r \ ,
\end{eqnarray}
where $\tau$ and $\kappa$ are given by
where $\tau$ and $\kappa$ are given by
\begin{equation}
\label{tau}
\tau=q\alpha/a^2   , \  \ \ \kappa=\frac{\lambda^2}{4\tau}-\frac{qj}{2} \ ,
\end{equation}
with
\begin{equation}
\label{2.30.1}
C^{(3)}_j=\left\{\begin{array}{ccc} \frac{\lambda}{\sqrt{\tau}}\frac{1}{2q|j|+1} \
\ j> 0 \ .
\\ -\frac{\sqrt{\tau}}{\lambda}(2q|j|+1), \ j<0 \ . \end{array}\right. \
\end{equation}
\end{enumerate}

For the second and third models, the radial functions are given in terms of
Whittaker functions, $M_{\kappa,\nu}(z)$.

For massless fields and at large distance from the core, the behavior of the 
core-induced azimuthal current can be developed as follows: instead to
use the summation on the angular moment $j$ in \eqref{Azimu3}, we use $n=j-1/2$.
In this way, we shall use a new notation, $F^{(i)}_{n}
\equiv F^{(i)}_{j}$. Moreover, we change $n$ by $n-n_0$, 
being $n_0$ given in \eqref{alpha}. So from \eqref{Azimu3} we can write,
\begin{eqnarray}
\label{Azimu4}
\langle j^\phi (x)\rangle_{c}=\frac{eq}{\pi^2 r^4}\int^{\infty}_{0} d z z^2 
\sum_{n=-\infty}^\infty K_{\beta}(z)
K_{{\tilde{\beta}}}(z)F^{(i)}_{n-n_0}(z (a/r))  \ .
\end{eqnarray}
In the above expression we are using the notation:  
\begin{eqnarray}
\label{F-ratio1}
F^{(i)}_{n-n_0}(z (a/r))= \frac{I_{{\tilde{\beta}}}(z (a/r))-
\mbox{Im}[{\cal{V}}_{n-n_0}^{(i)}(iz, (a/r))]I_{\beta}(z (a/r))}{K_{{\tilde{\beta}}}(z (a/r))+
\mbox{Im}[{\cal{V}}_{n-n_0}^{(i)}(iz, (a/r))]K_{\beta}(z (a/r))} \  .
\end{eqnarray}
In \eqref{F-ratio1} the orders of Bessel functions are given by
\begin{eqnarray}
\label{beta}
\beta=q|n+1/2+\alpha_0|-\frac12\frac{|n+1/2+n_0|}{n+1/2+\alpha_0} \ , \nonumber\\
{\tilde{\beta}}=q|n+1/2+\alpha_0|+\frac12\frac{|n+1/2+n_0|}{n+1/2+\alpha_0} \ .
\end{eqnarray}

Expanding the integrand of \eqref{Azimu4} in powers of $a/r$, the dominant term is given by the smaller
power of this ratio. So we have two possibilities: for $\alpha_0>0$ ($0\leq\alpha_0<1/2$) this term
is given by $n=-1$, and for $\alpha_0<0$ ($-1/2<\alpha_0\leq 0$) this term is given for $n=0$. 

Now, using the expansions for the modified Bessel functions for small arguments \cite{Abramo},
the leader contributions are:
\begin{itemize}
\item For $\alpha_0>0$:
\begin{equation}
\label{b6}
F^{(i)}_{-1-n_0}\left(z\frac{a}{r}\right)\approx \frac{2}{\Gamma^2(\beta)}
\frac{1+i{\cal{V}}_{-1-n_0}^{(i)}(iz, a/r)\left(z\frac{a}{r}\right)
\left(\frac{az}{2r\beta}\right)}{\frac{\Gamma(1-\beta)}
{\Gamma(\beta)}-i{\cal{V}}_{-1-n_0}^{(i)}(iz, a/r)
\left(z\frac{R}{r}\right)\left(\frac{az}{2r}\right)\left(\frac{2r}{az}\right)^{2\beta}} \  .
\end{equation}
\item For $\alpha_{0}<0$, and
\begin{equation}
\label{b7}
F^{(i)}_{-n_0}\left(z\frac{a}{r}\right)\approx -\frac{2}{\beta\Gamma^2(\beta)}
\frac{1+i{\cal{V}}_{-n_0}^{(i)}(iz, a/r)\left(z\frac{R}{r}\right)\left(\frac{2r\beta}{az}\right)}
{\left(\frac{2r}{az}\right)^{2\beta}-i\frac{\Gamma(1-\beta)}
{\Gamma(\beta)}{\cal{V}}_{-n_0}^{(i)}(iz, a/r)\left(z\frac{R}{r}\right)\left(\frac{az}{2r}\right)}  \  .
\end{equation}
\end{itemize}

The next steps are the calculations of the dominants contribution for
the coefficient that contains all the information about the core, ${\cal{V}}_{-1-n_0}^{(i)}(iz, a/r)$ 
and ${\cal{V}}_{-n_0}^{(i)}(iz, a/r)$, for the three models. This can be done by explicit 
substitution of the radial functions, $R^{(i)}_1(iz, a/r)$ and $R^{(i)}_2(iz, a/r)$,
into \eqref{Z.Bessel}.  So, for massless fields, we find:
\begin{equation}
\label{b13}
\langle j^\phi (r) \rangle_{c}\approx -2\frac{|\alpha_0|}{\alpha_0}\frac{eq}{\pi^2 r^4}\frac{\beta-\chi^{(l)}}
{\left(\frac{2r}{a}\right)^{2\beta}\chi^{(l)}}\frac{\beta}{2\beta+1} \  ,
\end{equation}
where
\begin{equation}
\label{b4-a}
\beta=q\left(\frac{1}{2}-|\alpha_{0}|\right)+\frac{1}{2}  \
\end{equation}
and $\chi^{(l)}$ is a parameter depending on the specific model adopted for the
magnetic field, given bellow by:
\begin{equation}
\label{b12}
\chi^{(l)}=\left\{\begin{array}{ccc}
\nu=q|n_0-\frac12\frac{|\alpha_0|}{\alpha_0}|-\frac12\frac{|\alpha_0|}{\alpha_0}, \ \mbox{for the model ({\it{i}})}\\
q\alpha(q+1)\frac{M_{\frac{q}{2}\frac{|\alpha_0|}{\alpha_0},\nu}(2q\alpha)}
{M_{\frac{q}{2}\frac{|\alpha_0|}{\alpha_0},\nu+1}(2q\alpha)}, \ \mbox{for the model ({\it{ii}})}\\
\frac{\sqrt{q\alpha}}{2}(q+1)\frac{M_{-\frac{1}{2}\frac{|\alpha_0|}{\alpha_0}
\frac{q+1}{2},\frac{\nu}{2}}(\tau R^2)}{M_{-\frac{1}{2}\frac{|\alpha_0|}
{\alpha_0}\frac{q+1}{2},\frac{\nu}{2}}(\tau R^2)}, \ \mbox{for the model ({\it{iii}})} \ .
\end{array}\right.
\end{equation}	
On basis of this results we can say that for the three models considered, 
the core-induced azimuthal current density decays with, $\frac1{r^4(a/r)^{2\beta}}$,
for large distance from the tube. We have shown in \cite{Mello13} that 
the massless limit of the zero-thickness azimuthal current given in \eqref{jazimu},
decays with $1/r^4$. So, we conclude that for large distance from the core, the total
azimuthal current, \eqref{Azimuthal3}, is dominated by the zero-thickness 
contribution.

Now let us investigate the behavior of the core-induced azimuthal current density 
near the core, for the three models. In general the current diverges in this region.
The dominant contribution in \eqref{Azimu3} comes from the contribution of large $|j|$.
To find the leading term it is convenient to introduce a new variable $z=\beta_j x$, and 
use the uniform expansion for large order for the modified Bessel functions \cite{Abramo}.
However, before to do that, we would like to notice that, 
changing $n\rightarrow -n-1$ the summation over \textit{j} keeps 
unchanged, but the parameter $\nu_j$ change as 
$\nu_j\rightarrow\tilde{\nu}_j$ and $\tilde{\nu}_j\rightarrow\nu_j$. If, in addition,
we also change $\alpha\rightarrow -\alpha$, then $\beta_j\rightarrow\tilde{\beta}_j$ 
and $\tilde{\beta}_j\rightarrow\beta_j$.  It means that, when we make $n\rightarrow -n-1$ 
and $\alpha\rightarrow -\alpha$ we have $ F^{(i)}_{j}(y)\longrightarrow- F^{(i)}_{j}(y)$.
On basis of this analysis, and considering $\alpha>0$, the behavior of the
core-induced azimuthal current near the boundary is given by, 
\begin{eqnarray}
\label{c4}
\langle j^\phi (x) \rangle_{c}\approx -2\frac{eq}{\pi^2 r^4}\sum_{n>0}\beta^{3}_{j}
\int^{\infty}_{\frac{mr}{\beta_j}}dx \ x^2 F^{(i)}_{j}(\beta_j x({a}/{r}))
K_{\beta_j}(\beta_j x)K_{\beta_j + \epsilon_j}(\beta_j x) \ .
\end{eqnarray}
Because we are considering $n>>1$, from now on we use the approximation, $\beta_j\approx\nu_j\approx qn$.

For the three models, it is necessary to find the leading term of $F^{(i)}_j$ for
large value of $j$. For the first two models, this term is obtained
by using the uniform expansion for large order for the modified Bessel functions
for the first model and the asymptotic expansion for the corresponding Whittaker 
functions for the second model. After some intermediate and long steps, we found
that both leading terms coincide and are given below,
\begin{eqnarray}
\label{c6}
F^{(1,2)}_{j}(qn x(a/r))\approx \frac{1}{4q^2\pi}\frac{e^{2qn\tilde{\eta}}}{ n^2 (1+e^{2\tilde{\eta}})} \  ,
\end{eqnarray}
where $\tilde{\eta}=\sqrt{1+x^2(a/r)^2}$. The third model converge to the same result. In order to see
that we express the Whittaker function in terms of the confluent hypergeometric functions \cite{Abramo}, 
$M_{\kappa,\mu}(z)=e^{-z/2}z^{1/2+\mu}M(1/2+\mu-\kappa,1+2\mu;z)$, and use the expansion below for the 
latter,
\begin{eqnarray}
\label{c5}
M(a,b;z)=\Gamma(b)e^{zx}\sum^{\infty}_{k=0} C_k z^k (-az)^{\frac{1}{2}(1-b-k)} J_{b-1+k}(2\sqrt{-az}) \ .
\end{eqnarray}
Finally we use the uniform expansion for large order for the Bessel function. So, for the 
three models, the core-induced azimuthal current density is given by
\begin{eqnarray}
\label{c7}
\langle j^\phi(x) \rangle_c \approx -\frac{eq}{4\pi^2 r^4}\sum_{n>0}\int^{\infty}_{\frac{mr}{qn}}
dz \ z^2\frac{e^{-2qn(\eta-\tilde{\eta})}}{(1+e^{2\tilde{\eta}})\sqrt{1+z^2}} \ ,
\end{eqnarray}
where $\eta=\sqrt{1+x^2}$. Using the approximation $\eta-\tilde{\eta}\approx z(1-a/r)$, and 
observing that the denominator of the integrand in \eqref{c7} can be approximate to unity, for the 
three models, we have:
\begin{eqnarray}
\label{c8}
\langle j^\phi(x)\rangle_c \approx -\frac{eq}{4\pi^2 r^4}\sum_{n>0} \int^{\infty}_{\frac{mr}{qn}}
dz \ z^2 e^{-2qn(1-a/r)}\ .
\end{eqnarray}
Solving the above integral we have
\begin{eqnarray}
\label{c9}
\langle j^\phi(r) \rangle_c \approx -\frac{e}{(4\pi q )^2}\frac1 r\frac1{(r-a)^3} \  . 
\end{eqnarray}
Here, we must notice that the current density diverges near the boundary. 

Because the zero-thickness azimuthal current, $\langle j^\phi(r) \rangle_s$, presents a finite value
near the boundary, we can conclude that, in this region, the total azimuthal current, 
\eqref{Azimuthal3} is dominated by the core-induced contribution. 

An alternative way to show that the three models provide the same result for the core-induced azimuthal
current near the boundary, is given by analyzing the radial differential equations \eqref{a1} and 
\eqref{a2} given in Appendix \ref{AppendA} for large value of $j$. Replacing the parameter 
$\lambda$ by $i\nu_j z$, specifically for \eqref{a1}, we get:\footnote{For $R_2(r)$ the equation is
similar to \eqref{Eq-1} by changing $\nu_j$ by $\nu_j+\frac{\tilde{\epsilon}_j}{2}$.}
\begin{equation}
\label{Eq-1}
r^2 R_1''(r) +rR_1'(r)-r^2\nu^2_j z^2R_1(r)-\nu_j^2\left(1+\frac{2eA_\phi(r)}{\nu_j}\right)R_1(r)\approx 0 \ .
\end{equation} 
Substituting \eqref{AModel1}, \eqref{AModel2} or \eqref{AModel3} into the above equation, 
it is possible to see that the solutions can be written in a general form below:
\begin{equation}
R_1(r)=I_{\nu_j}(\nu_j z r)f(r) \ , 
\end{equation}
where $f(r)$ is equal to unity for the first model, Eq. \eqref{AModel1}, and 
for the second and third models, it reads,
\begin{equation}
f(r)\approx 1-{\frac {rq\alpha}{a}}+\frac{q\alpha z} 2\left( {\frac {q\alpha}{az}}+1
 \right) \frac{{r}^{2}}{{a}}+O \left( {\nu_j}^{-1} \right) 
\end{equation}
and
\begin{equation}
f(r)\approx {{\rm e}^{-{\frac {rq\alpha}{{a}^{2}z}}}} \left( 1+{\frac {rq\alpha}{{
a}^{2}z}}-\frac{q\alpha}2 \left(1 -{\frac {q\alpha}{{z}^{2}{a}^{2}}}
 \right)\frac{ {r}^{2}}{{a}^2} +O \left( {\nu_j}^{-1} \right)\right)  \  , 
\end{equation}
respectively. 

Because the coefficient ${\cal{V}}_j^{(i)}$ depends on the ratio of the two radial 
functions, $R_2(r)$ and $R_1(r)$, for large value of $j$, the leading term of this ration is:
\begin{equation}
\frac{R_2(r)}{R_1(r)}\approx \frac{I_{\nu_j+{\tilde{\epsilon}}_j/2}((\nu_j+{\tilde{\epsilon}}_j/2)z r)}
{I_{\nu_j}(\nu_j z r)} \ . 
\end{equation}
Consequently the three models behave, in this limit, in a similar way.

After these above general descriptions of the core-induced azimuthal current, 
we would like to provide additional informations which are not easily 
observed by the analytical expressions for it. So, in order to full fill  
this lack, in the rest of this section we will develop some numerical evaluations.

\begin{figure}[!htb]
\begin{center}
\includegraphics[width=0.4\textwidth]{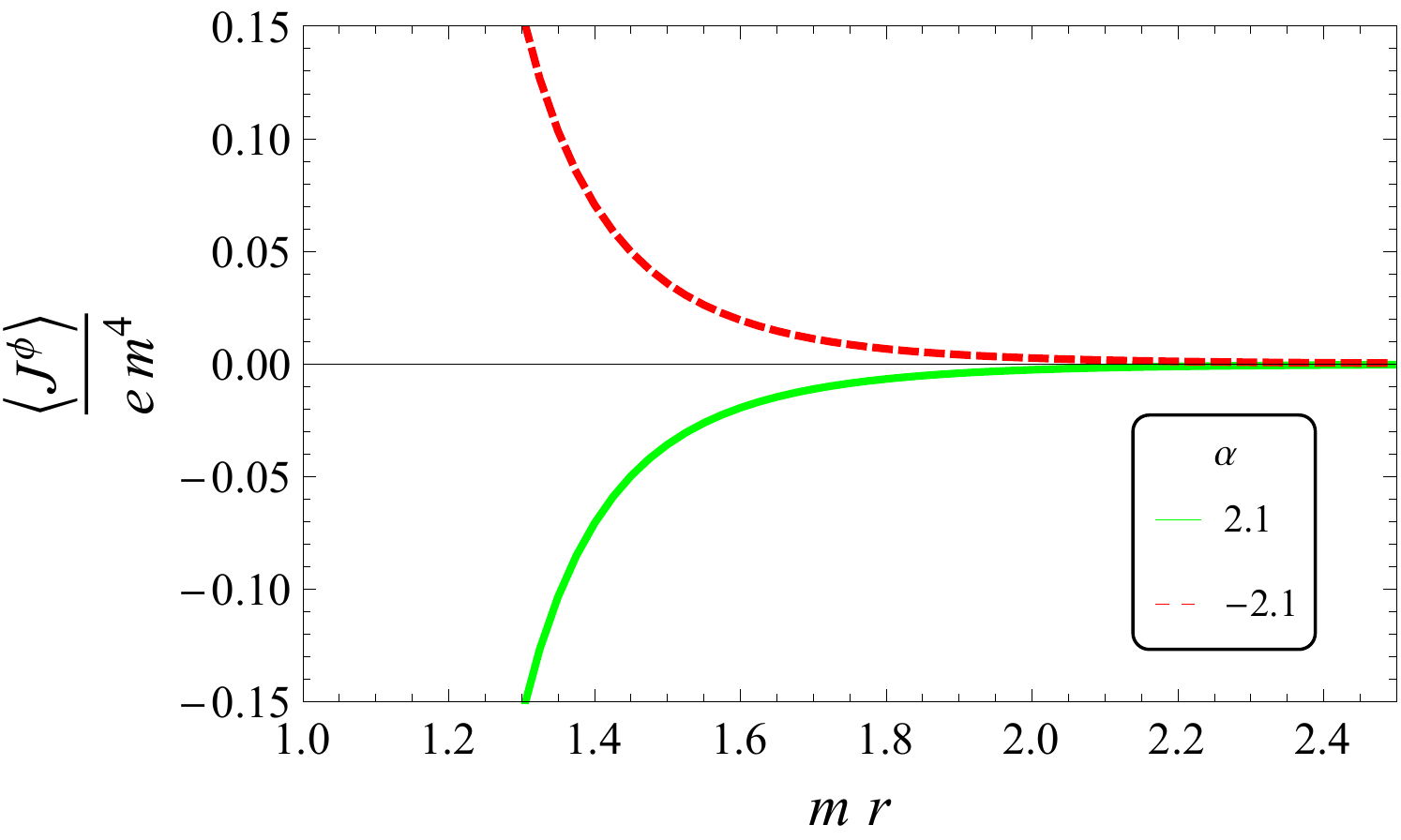}
\includegraphics[width=0.4\textwidth]{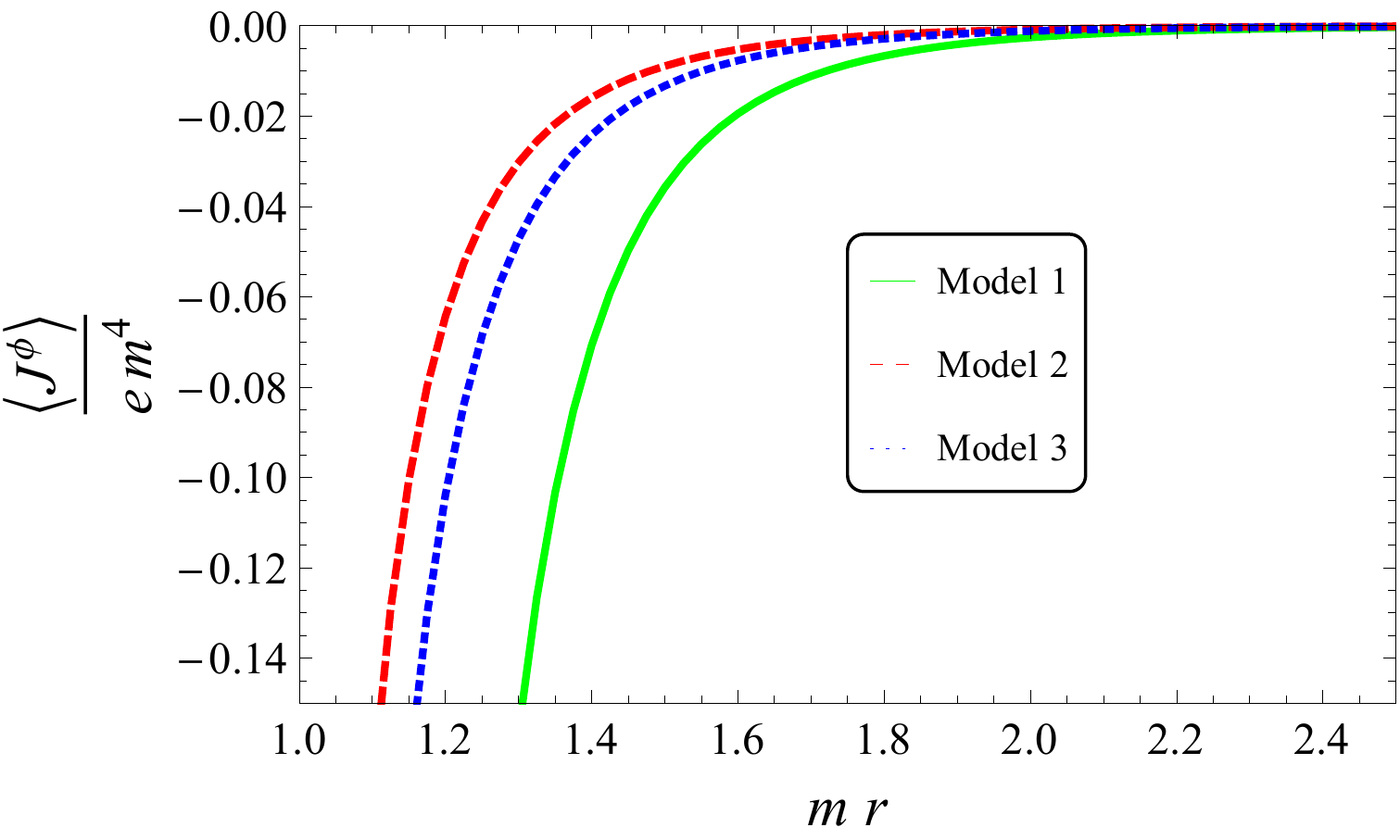}
\caption{The core-induced azimuthal current density is plotted,
in units of ``$m^4e$", as function of $mr$ for the  values $q=1.5$ and
$ma=1$. In the left plot we consider the current induced by the magnetic 
field configuration of the first model, and taking $\alpha=2.1$ and $\alpha=-2.1$.
On the right plot we compare the intensity of the core-induced current
for the three different models of magnetic fields considering $\alpha=2.1$.}
\label{fig1}
\end{center}
\end{figure}

In Fig. \ref{fig1} we display the dependence of the core-induced azimuthal current 
density, $\langle j^\phi(r) \rangle_c$, as function of $mr$ considering $q=1.5$ and
$ma=1$. On the left plot, we present the behavior for the current induced by the cylindrical
shell of magnetic field, taking into account positive and negative value for $\alpha$.
In order to provide a better understanding about this current, on the
right plot we exhibit its behavior as function of $mr$, 
for the three different models of magnetic fields considering 
$\alpha=2.1$. By this plot we can infer that for a given 
point outside the tube, the intensity of the current induced by the first model is 
the biggest one. 

\begin{figure}[!htb]
\begin{center}
\includegraphics[width=0.4\textwidth]{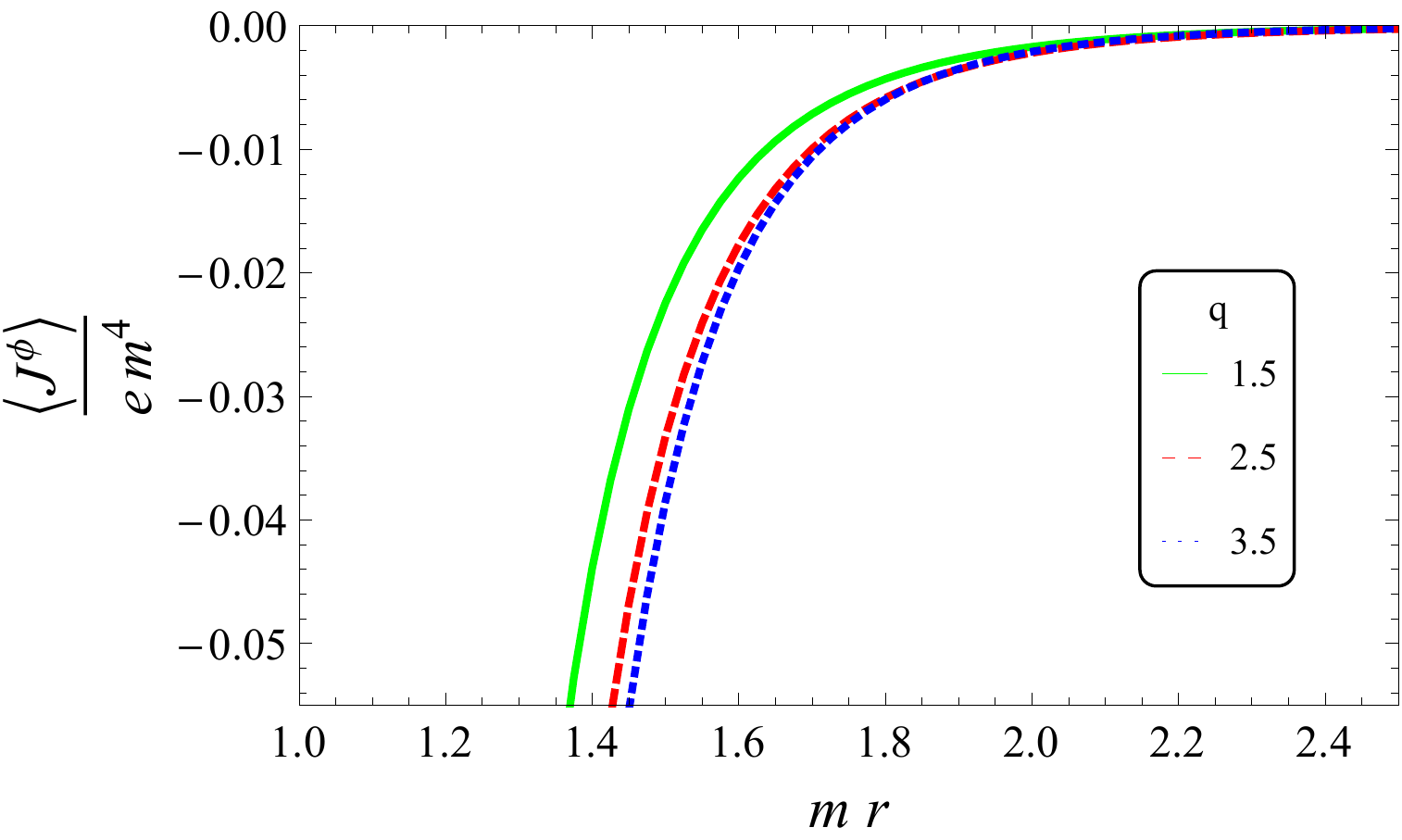}.
\caption{The core-induced azimuthal current density in units of ``$m^4e$", 
due to a magnetic field concentrated on a shell, as function of $mr$ 
considering three different values of $q$. In this plot we adopted $\alpha=1.2$.}
\label{fig2}
\end{center}
\end{figure}

Another analysis that deserves to be developed is related with dependence of the
core-induced azimuthal current with the parameter which codifies the presence 
of the cosmic string, $q$. So, in Fig \ref{fig2}, we display,
for the magnetic field concentrated in a cylindrical shell, the behavior of
 $\langle j^\phi(r) \rangle_c$ as function of $mr$ for 
$q=1.5, \ 2.5, \ 3.5$. As we can see the intensity of the current increases with $q$.
For this plot we consider $\alpha=1.2$ and $ma=1$.

\begin{figure}[!htb]
\begin{center}
\includegraphics[width=0.4\textwidth]{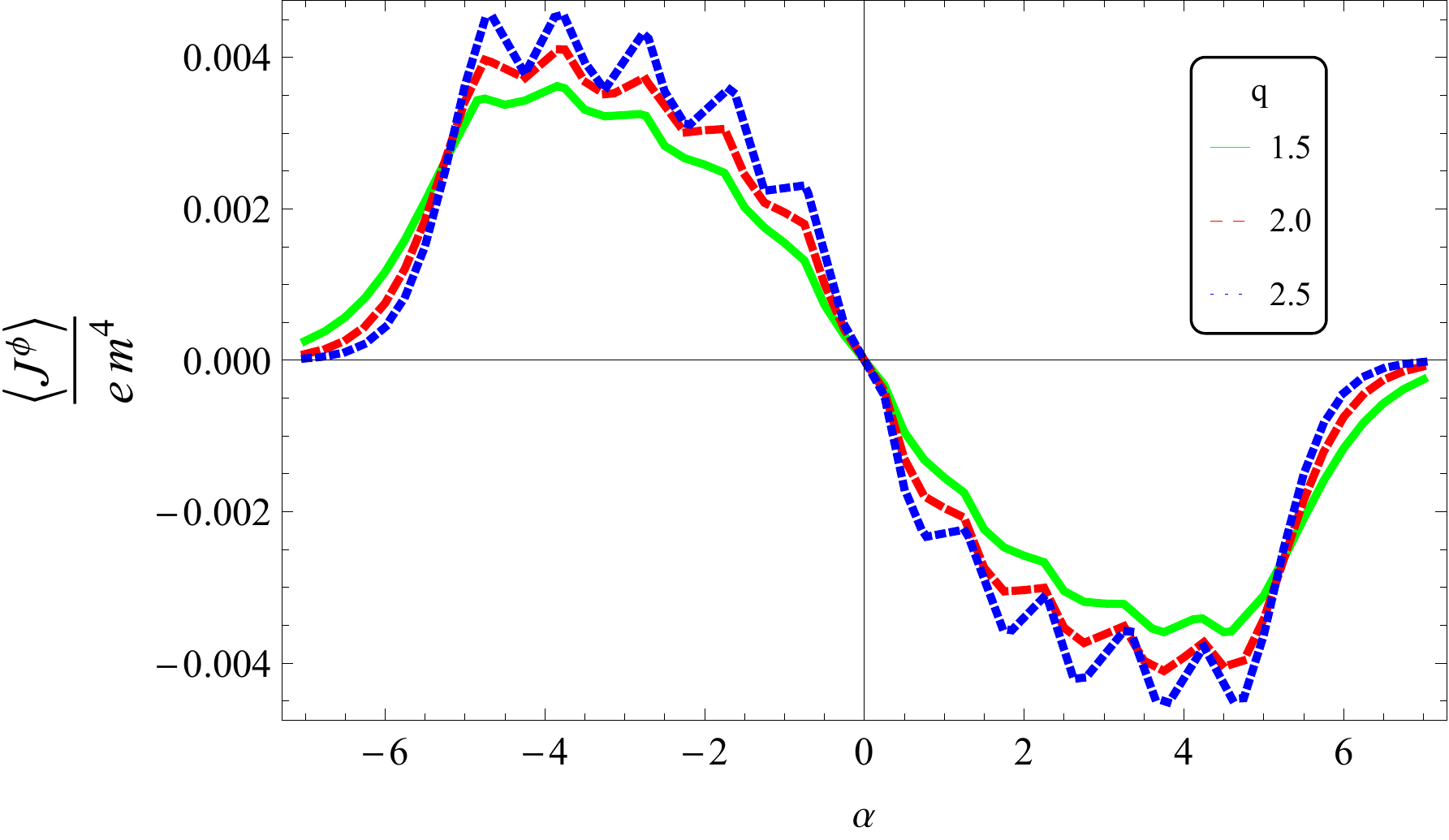}
\includegraphics[width=0.4\textwidth]{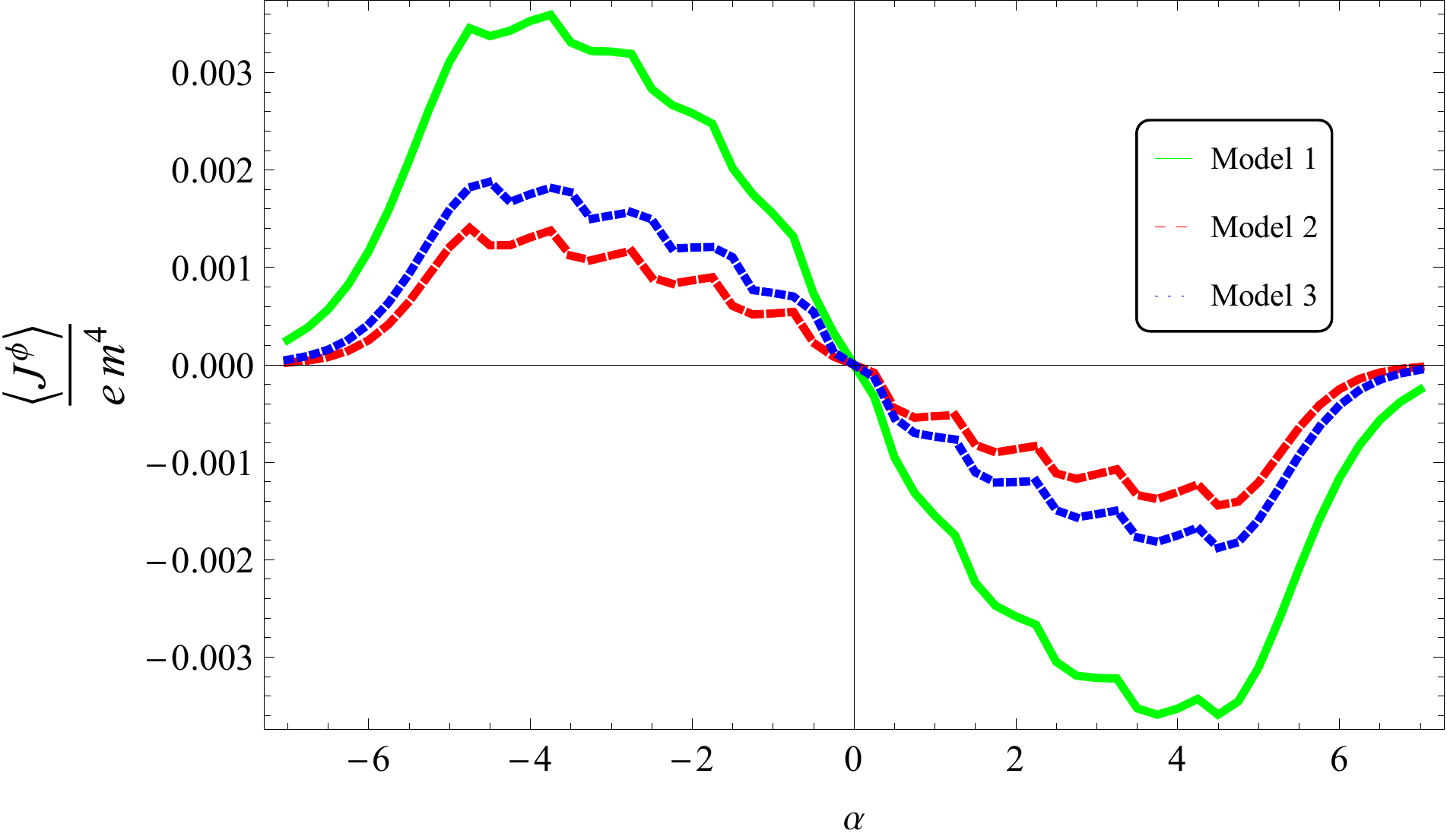}
\caption{The core-induced azimuthal current density is plotted,
in units of ``$m^4e$", as function of $\alpha$ for $ma=1$ and $mr=2$.
In the left plot, we exhibit only the current induced by the
first model considering different values of $q$, and in the 
right plot, we exhibit the current induced by the three different models 
of magnetic fields, considering $q=1.5$.}
\label{fig3}
\end{center}
\end{figure}

Finally in Fig \ref{fig3} we exhibit the behavior of the core-induced current
as function of $\alpha$, considering $ma=1$ and $mr=2$. In the left plot, we display 
the current induced by the first model of magnetic field and different 
values of $q$. These values are $q=1.5, \ 2.0, \ 3.5$.
In the right plot we consider the currents induced by the three different models,
adopting $q=1.5$. For both plots, we assume that $\alpha$ varies in the
interval $[-7.0, \ 7.0]$. From both plots we can infer once more that,
the intensity of the current increases with $q$ (left plot) and the first model provide the
current with biggest intensity (right plot).

\section{Conclusions}
\label{Conclu}

In this paper we have investigated the influence of the conical topology of the spacetime 
and the finite core effect on the vacuum expectation value of fermionic current densities
induced by magnetic fluxes. In the model considered we have adopted that
the geometry of the spacetime corresponds to an idealized cosmic string everywhere, surrounded 
by a magnetic tube of radius $a$. In this tube three different configurations of
magnetic fields have been taken into account: a cylindrical shell, a field decaying 
with $1/r$ and finally a homogeneous magnetic field. In order to develop these 
analysis we had to construct the normalized fermionic wave-function for the 
region outside the tube, and calculated the fermionic current densities 
by using the mode summation method, as shown in \eqref{current} and \eqref{sum}. 
This complete set of the fermionic wave-function was constructed by imposing 
continuity condition for the wave-functions in the 
regions inside and outside the tube, at the boundary, $r=a$. These functions were
given by \eqref{in} and \eqref{psi-out}, respectively. As to the normalization condition, it
determines the values of the interior radial functions at the boundary, see \eqref{Rela4}.
 
Although the magnetic field vanishes outside the tube, the magnetic field inside the tube induced a non-vanishing
only an azimuthal current density in the exterior region. This outside
current is consequence of two type of effects: the Aharonov-Bohm-like effect, and
the direct interaction of the charged fermionic field with the magnetic field inside the core.
This phenomenon is explicitly manifested in the structure of the induced current.
The latter been decomposed in two distinct contributions:
The first one depends only on the fractional part of the ratio of the
total magnetic flux by the quantum one. It is given in \eqref{jazimu}
and is a periodic function of the total flux with the period equal to quantum flux. 
The second contribution named core-induced given by \eqref{Azimu3}, in general
is not a periodic function of the magnetic flux and depends on the total magnetic 
flux inside the core.  By general analysis we could observe that the
core-induced azimuthal current decays with $e^{-2mr}/(mr)^3$
for $mr>>1$ (see \eqref{massive4}). For the zero-thickness 
azimuthal current the corresponding decay is $e^{-2mr\sin(\pi/q)}/(mr)^{5/2}$
for $q>2$, consequently the latter dominates the 
total azimuthal current. For $q\leqslant 2$, both
contributions are of the same order. We have also analyzed the core-induced current
for two other asymptotic regions of the parameters. In these analysis
we have explicitly used the solutions for the radial function inside the tube
for the three different models of magnetic flux. 
\footnote{The radial solutions for the region inside the core, were possible by 
the obtainment of recurrence relations involving Whittaker 
function $M_{\kappa,\lambda}(z)$ given in the Appendix \ref{AppendA}.}
For massless field and at large distance from the core, the result
was given by Eq. \eqref{b13}, and explicitly we see that this current decays with
$\frac1{r^4(a/r)^{2\beta}}$ for large distance from the tube.
Comparing this result with the corresponding one for 
the massless limit of the zero-thickness azimuthal current
wiche decays with $1/r^4$, we conclude that for large distance the total azimuthal
current, is dominated by the zero-thickness contribution.
For point near the tube core, the core-induced current 
diverges with $\frac1{r^4(1-a/r)^3}$, as shown by \eqref{c9}. So this contribution
is dominant in this region. 

Finally we have also provide, by using numerical evaluation, the
behavior of the core-induced current as function of several 
physical quantities relevant in our analysis. In figure 1 we have two plots.
In the left plot an expected result is presented. The current changes its sign
when we change the sign of $\alpha$. In the right plot, it is exhibited
the behavior of the current for the three models as function of $mr$. It
is shown that the intensity of the current induced by the first model is the biggest one.
In figure 2, we exhibit the behavior of current density for the
first model as function of $mr$ and fixed value of $\alpha$ for different
values of $q$. By this plot we can infer that increasing $q$
the intensity of the current also increases. (Although this latter numerical
analysis was presented only for the first model, for the other two models
the current behaves in a similar way.) In figure 3, we have displayed  
current as function of the intensity of the magnetic flux. In the left
plot we have considered only the first model fixed $mr$ and 
varying $q$. This plot reinforces the fact that the intensity of the current
increases with $q$; moreover, it shows that the current is not a periodic 
function of the flux. The right plot exhibits the behavior of the
core-induced current, for the three models, for fixed value of $q$. 
Also this plot reinforce that the first model provides the current with 
biggest intensity.\footnote{In all numerical analysis it was consideres $ma=1$.}

\textbf{Acknowledgments}: We thank Coordena\c{c}\~ao de Aperfei\c{c}oamento 
de Pessoal de N\'{\i}vel Superior (CAPES) and Conselho Nacional de Desenvolvimento 
Cient\'{\i}fico e Tecnol\'ogico (CNPq) for partial financial support.

\appendix

\section {Recurrence relations for the radial}
\label{AppendA}

In this appendix we shall develop explicitly the radial differential equations
satisfied by functions $R_1(r)$ and $R_2(r)$, considering the three
configurations of magnetic fields given by the ansatz \eqref{2.10} in (\ref{2.11}):
\begin{eqnarray}
\label{a1}
\left[r^2\lambda^2-\left(qj+eA_\phi^{(i)} - \frac{1}{2}\right)^2 - 
er^2B^{(i)}(r)\right]R_1(r) +r^2R^{''}_1(r)+rR^{'}_1(r)=0 \  ,
\end{eqnarray}
\begin{eqnarray}
\label{a2}
\left[r^2\lambda^2-\left(qj+eA^{(i)}_\phi + \frac{1}{2}\right)^2 + 
er^2B^{(i)}(r)\right]R_2(r)+r^2R^{''}_2(r)+rR^{'}_2(r)=0  \  ,
\end{eqnarray}
where the upper index $i=1, \ 2, \ 3$ characterizes which model we are working.

Now we have the general system of equation to the inside situation 
for each kind of magnetic field that we are working the above equation will 
provide different types of solution to the radial functions.

\subsection{The cylindrical shell of magnetic field}
\label{SA1}

For this case we have for the vector potential
\begin{eqnarray}
\label{AModel1}
A^{(1)}_\phi=-q\frac{\Phi}{2\pi}\Theta(r-a) \ .
\end{eqnarray}
For this case, in the the region $r<a$, both vector potential
and magnetic field vanish. So the equations \eqref{a1} and \eqref{a2} reads,
\begin{eqnarray}
\label{a01}
\left[r^2\lambda^2-\left(qj - \frac{1}{2}\right)^2\right]R_1(r)+r^2 R''_1(r)+rR'_1(r)=0 \  ,
\end{eqnarray}
\begin{eqnarray}
\label{a02}
\left[r^2\lambda^2-\left(qj + \frac{1}{2}\right)^2\right]R_2(r)+r^2R''_2(r)+rR'_2(r)=0  \  .
\end{eqnarray}

The regular solutions at origin, $r=0$, are the first kind Bessel functions. So, for the upper components
of the fermionic wave-function we have
\begin{equation}
\label{a03}
\psi_{+}=e^{ikz}e^{iqn\phi}\left( \begin{array}{cc}
a J_{\nu_j}(\lambda r) \\
b J_{\nu_j+\tilde{\epsilon}_j}(\lambda r)e^{iq\phi} \end{array} \right) \ ,
\end{equation}
where $\nu_j=q|j|-\frac{\tilde{\epsilon}_j}{2}$, with $\tilde{\epsilon}_j=1$ for $j>0$ and
$\tilde{\epsilon}_j=-1$ for $j<0$.
The lower component is found when we substitute the above solution into (\ref{2.11}), then we find
\begin{equation}
\label{a04}
\psi_{-}=\frac{-ie^{ikz}e^{iqn\phi}}{(E+m)}\left( \begin{array}{cc}
\lambda b [J^{'}_{\nu_j+\tilde{\epsilon}_j}(\lambda r)+\frac{\nu_j +
\epsilon_j}{\lambda r}J_{\nu_j+\tilde{\epsilon}_j}(\lambda r)]+ik a J_{\nu_j}(\lambda r) \\
\{\lambda a [J^{'}_{\nu_j}(\lambda r)-\frac{\nu_j}{\lambda r}J_{\nu_j}(\lambda r)]-
ik b J_{\nu_j+\tilde{\epsilon}_j}(\lambda r)\}e^{iq\phi} \end{array} \right) \  .
\end{equation}
By using the recurrence relations involving the Bessel functions \cite{Abramo}, we can write:
\begin{equation}
\label{a05}
\psi_{-}=e^{ikz}e^{iqn\phi}\left( \begin{array}{c}
c J_{\nu_j}(\lambda r) \\
d J_{\nu_j+\tilde{\epsilon}_j}(\lambda r)e^{iq\phi} \end{array} \right),
\end{equation}
where the new coefficients $c$ and $d$ are given below,
\begin{eqnarray}
\label{a06}
c=\frac{ka-i\epsilon_j\lambda b}{E+m}
\end{eqnarray}
\begin{eqnarray}
\label{a07}
d=-\frac{kb-i\epsilon_j\lambda a}{E+m}
\end{eqnarray}

\subsection{The magnetic field decays with $1/r$}
\label{SA2}

For this case we have the vector potential given by
\begin{eqnarray}
\label{AModel2}
A^{(2)}_\phi=-q\frac{\Phi r}{2\pi a} \ .
\end{eqnarray}
The magnetic field is $B^{(2)}=-q\frac{\Phi }{2\pi ar}$. So by (\ref{a1}) and (\ref{a2}) follow that,
\begin{equation}
\label{a1.1}
\left[r^2\lambda^2-\left(qj+q\alpha\frac{r}{a} - \frac{1}{2}\right)^2 -
\frac{q\alpha r}{a}\right]R_1(r)+r^2 R''_1(r)+rR'_1(r)=0 \  ,
\end{equation}
\begin{equation}
\label{a2.2}
\left[r^2\lambda^2-\left(qj+q\alpha\frac{r}{a} + \frac{1}{2}\right)^2 +
\frac{q\alpha r}{a}\right]R_2(r)+r^2R''_2(r)+rR'_2(r)=0 \  .
\end{equation}

The regular solutions at origin of above equations are the Whittaker 
function $M_{\kappa,\lambda}(z)$.
So, for the upper component we have
\begin{equation}
\label{a3}
\psi_{+}=\frac{e^{ikz}e^{iqn\phi}}{\sqrt{r}}\left( \begin{array}{cc}
c_{1}M_{\kappa,\nu_j}(\xi r) \\
c_{2}M_{\kappa,\nu_j+\tilde{\epsilon}_j}(\xi r)e^{iq\phi} \end{array} \right),
\end{equation}
where the parameters $\kappa$ and $\xi$ are given in (\ref{2.300}). The lower component
is found when we substitute the above solution into (\ref{2.11}). We have:
\begin{eqnarray}
\label{a4}
\psi_{-}=\frac{-ie^{ikz}e^{iqn\phi}}{(E+m)\sqrt{r}}\left( \begin{array}{cc}
\xi c_{2}[M^{'}_{\kappa,\nu_j+\tilde{\epsilon}_j}(\xi r)+\frac{qj+
\alpha r/a}{\xi r}M_{\kappa,\nu_j+\tilde{\epsilon}_j}(\xi r)]+ikc_1 M_{\kappa,\nu_j}(\xi r) \\
\{\xi c_{1}[M^{'}_{\kappa,\nu_j}(\xi r)-\frac{qj+\alpha r/a}{\xi r}M_{\kappa,\nu_j}(\xi r)]-
ikc_2 M_{\kappa,\nu_j+\tilde{\epsilon}_j}(\xi r)\}e^{iq\phi} \end{array} \right) \  .
\end{eqnarray}

Unfortunately, we did not find in the literature any recurrence relation
involving this kind of Whittaker functions which allows us to express
the lower components in a simpler form. So, by expressing these
functions in terms of Confluent Hypergeometric Functions, and by
using the computer program MAPLE, we could construct the following
relations:
\begin{itemize}
\item For $j>0$, we have,
\begin{eqnarray}
\label{a5.a}
M'_{\kappa,\nu_j+1}(z)+\left(\frac{\nu_j+1/2}{z}
-\frac{\kappa}{2\nu_j + 1}\right)M_{\kappa,\nu_j + 1}(z)=c^{(+)}_j M_{\kappa,\nu_j}(z) \ ,
\end{eqnarray}
\begin{eqnarray}
\label{a6.a}
M'_{\kappa,\nu_j}(z)-\left(\frac{\nu_j+1/2}{z}
-\frac{\kappa}{2\nu_j + 1}\right)M_{\kappa,\nu_j}(z)=c^{(-)}_j M_{\kappa,\nu_j+1}(z) \ .
\end{eqnarray}
\item For $j<0$, we have:
\begin{eqnarray}
\label{a5.b}
M'_{\kappa,\nu_j-1}(z)-\left(\frac{\nu_j-1/2}{z}
-\frac{\kappa}{2\nu_j - 1}\right)M_{\kappa,\nu_j - 1}(z)=c^{(+)}_j M_{\kappa,\nu_j}(z) \ ,
\end{eqnarray}
\begin{eqnarray}
\label{a6.b}
M'_{\kappa,\nu_j}(z)+\left(\frac{\nu_j-1/2}{z}
-\frac{\kappa}{2\nu_j - 1}\right)M_{\kappa,\nu_j}(z)=c^{(-)}_j M_{\kappa,\nu_j-1}(z) \  .
\end{eqnarray}
\end{itemize}

In the above expressions the coefficients $c_j^{(\pm)}$ are given below:
\begin{eqnarray}
\label{a7}
c^{(+)}_j&=&\left\{\begin{array}{cc} 2(\nu_j + 1) \ , \  j> 0 \ . \\
\frac{1}{8\nu_j}\left[1-\left(\frac{2\kappa}{2\nu_j + 1}\right)^2\right]
\ , \  j<0 \  .
\end{array}
\right.\\
\label{a8}
c^{(-)}_j&=&\left\{\begin{array}{cc}
-\frac{1}{8(\nu_j + 1)}\left[1-\left(\frac{2\kappa}{2\nu_j + 1}\right)^2\right]\ , \  j>0 \ . \\
2\nu_j\ , \ j<0 \  .
\end{array}\right.
\end{eqnarray}
Finally we can write the Dirac wave-function with positive-energy into the form below
\begin{equation}
\label{a9}
\psi^{(+)}(x)=\frac{e^{-ipx}e^{iqn\phi}}{\sqrt{r}}\left( \begin{array}{c}
c_{1}M_{\kappa,\nu_j}(\xi r) \\
c_{2}M_{\kappa,\nu_j+\tilde{\epsilon}_j}(\xi r)e^{iq\phi}\\
\tilde{c}_{1}M_{\kappa,\nu_j}(\xi r)\\
\tilde{c}_{2}M_{\kappa,\nu_j+\tilde{\epsilon}_j}(\xi r)e^{iq\phi} \end{array} \right),
\end{equation}
where
\begin{eqnarray}
\label{a10}
\tilde{c}_{1}=\frac{kc_{1}-ic_{2}c^{(+)}_j}{E+m} \, \ \ \ \tilde{c}_{2}=-\frac{kc_{2}+ic_{1}c^{(-)}_j}{E+m} \  . 
\end{eqnarray}

\subsection{The homogeneous magnetic field}
\label{A3.3}

Our last analysis is for uniform magnetic field. In this case the azimuthal
component of the vector potential is,
\begin{eqnarray}
\label{AModel3}
A^{(3)}_\phi(r)=-\frac{q\Phi}{2\pi}\left(\frac{r}{a}\right)^2 \ .
\end{eqnarray}

The magnetic fields is $B^{(3)}=-\frac{2q\Phi}{2\pi a^2}$.
Following the same proceeding as before, the solutions will be also
given in terms of the regular at origin Whittaker functions.
For the upper component we have
\begin{equation}
\label{upper}
\psi_{+}=\frac{e^{ikz}e^{iqn\phi}}{r}\left( \begin{array}{cc}
c_{1}M_{\kappa-\frac{1}{4},\frac{\nu_j}{2}}(\tau r^2) \\
c_{2}M_{\kappa+\frac{1}{4},\frac{\nu_j+\tilde{\epsilon}_j}{2}}(\tau r^2)e^{iq\phi} \end{array} \right) \ ,
\end{equation}
where $\tau$ and $\kappa$ are given by
\begin{equation}
\label{30.032}
\tau=q\alpha/a^2   , \  \ \ \kappa=\frac{\lambda^2}{4\tau}-\frac{qj}{2} \ .
\end{equation}

Substituting \eqref{upper} into (\ref{2.11}) we have the lower component given by
\begin{eqnarray}
\label{lower}
\psi_{-}=\frac{-ie^{ikz}e^{iqn\phi}}{(E+m)r}\left( \begin{array}{cc}
c_{2}\left(\frac{qj-\frac{1}{2}}{r}+\tau r +\partial_r\right)
M_{\kappa^+,\frac{\tilde{\nu}_j}{2}}(\tau r)+ikc_1 M_{\kappa^-,\frac{\nu_j}{2}}(\tau r) \\
\{c_{1}\left(\partial_r-\frac{qj-\frac{1}{2}+\tau r^2}{r}\right)
M_{\kappa^-,\frac{\nu_j}{2}}(\tau r)-ikc_2 M_{\kappa^+,\frac{\tilde{\nu}_j}{2}}(\tau r)\}e^{iq\phi} \end{array} \right).
\end{eqnarray}

 Also we have not found in literature the recurrence relations involving the Whittaker 
functions necessary for us to express the \eqref{lower} in a simpler form. 
So, by using the same procedure explained in the last analysis we
were able to find the following recurrence relations:
\begin{itemize}
\item For $j>0$:
\begin{eqnarray}
\label{a12.1}
\left(\frac{\nu_j }{2\tau}+\frac{1}{2} +\frac{d}{dz}\right)M_{\kappa+\frac{1}{4},\frac{\nu_j+1}{2}}(z)
=\frac{c^{(+)}_{j}}{2\sqrt{z}} M_{\kappa-\frac{1}{4},\frac{\nu_j}{2}}(z) \  ,
\end{eqnarray}
\begin{eqnarray}
\label{a12.2}
\left(-\frac{\nu_j }{2\tau}+\frac{1}{2} +\frac{d}{dz}\right)M_{\kappa-\frac{1}{4},\frac{\nu_j}{2}}(z)
=\frac{c^{(-)}_{j}}{2\sqrt{z}} M_{\kappa+\frac{1}{4},\frac{\nu_j+1}{2}}(z)\  .
\end{eqnarray}
\item For $j<0$
\begin{eqnarray}
\label{a12.3}
\left(-\frac{\nu_j + 1}{2\tau}+\frac{1}{2} +\frac{d}{dz}\right)M_{\kappa+\frac{1}{4},\frac{\nu_j - 1}{2}}(z)
=\frac{c^{(+)}_{j}}{2\sqrt{z}} M_{\kappa-\frac{1}{4},\frac{\nu_j}{2}}(z) \  ,
\end{eqnarray}
\begin{eqnarray}
\label{a12.4}
\left(\frac{\nu_j + 1}{2\tau}+\frac{1}{2} +\frac{d}{dz}\right)M_{\kappa-\frac{1}{4},\frac{\nu_j}{2}}(z)
=\frac{c^{(-)}_{j}}{2\sqrt{z}} M_{\kappa+\frac{1}{4},\frac{\nu_j-1}{2}}(z) \ .
\end{eqnarray}
\end{itemize}
In the above expressions, $c_j^{(\pm)}$ are given by,
\begin{eqnarray}
\label{a13}
c^{(+)}_j&=&\left\{\begin{array}{cc} 2(\nu_j+1), \ j>0 \\
 \left(1-\frac{4\kappa+1}{2(\nu_j+1)}\right), \ j< 0 \end{array}\right. \\
\label{a14}
c^{(-)}_j&=&\left\{\begin{array}{cc} -\left(1+\frac{4\kappa-1}{2\nu_j}\right), \ j>0 \\
 2\nu_j, \ j< 0\end{array}\right.
\end{eqnarray}

Finally for the third model we have the positive-energy wave-function written below
\begin{equation}
\label{a15}
\psi^{(+)}(x)=\frac{e^{-ipx}e^{iqn\phi}}{r}\left( \begin{array}{c}
c_{1}M_{\kappa^-,\frac{\nu_j}{2}}(\tau r^2) \\
c_{2}M_{\kappa^+,\frac{\tilde{\nu}_j}{2}}(\tau r^2)e^{iq\phi} \\
\tilde{c}_{1}M_{\kappa^-,\frac{\nu_j}{2}}(\tau r^2)\\
\tilde{c}_{2}M_{\kappa^+,\frac{\tilde{\nu}_j}{2}}(\tau r^2)e^{iq\phi} \end{array} \right),
\end{equation}
where $\tilde{c}_{1,2}$ are also given by (\ref{a10}).

\subsection{Analysis of the coefficient ${\mathcal{V}}^{(i)}_{j}$}

To prove \eqref{Rel-V} we shall use its explicit form given bellow:
\begin{eqnarray}
\label{ratioprove}
{\mathcal{V}}^{(i)}_{j} (\pm i \lambda, a )=\frac{R^{(i)}_{2} (r)}{R^{(i)}_{1}(r)} \ .
\end{eqnarray}
For the first model we have
\begin{eqnarray}
\label{ratioprove1}
{\mathcal{V}}^{(1)}_{j}(\pm i \lambda , a)=\tilde{\epsilon}_j \frac{J_{\tilde{\nu}_j}(i\lambda a)}{J_{\nu_j}(i\lambda a)} \ .
\end{eqnarray}
Using the well know relation involving the Bessel function with imaginary 
argument and the modified Bessel function \cite{Abramo}, the above equation becomes 
\begin{eqnarray}
\label{ratioprove2}
{\mathcal{V}}^{(1)}_{j} (\pm i \lambda , a)=\tilde{\epsilon}_j (\pm i)^{\tilde{\epsilon}_j} 
\frac{I_{\tilde{\nu}_j}(\lambda a)}{I_{\nu_j}(\lambda a)} \ .
\end{eqnarray}
We notice that, for positive or negatives values of $j$, $\tilde{\epsilon}_j=\pm 1$, 
consequently $\tilde{\epsilon}_j (\pm i)^{\tilde{\epsilon}_j}= \pm i$. In this way we have:
\begin{eqnarray}
\label{ratioprove3}
{\mathcal{V}}^{(1)}_{j} (\pm i \lambda , a)=\pm i{\rm Im}[\mathcal{V}^{(1)}_{j}(i\lambda, a)].
\end{eqnarray}

For the others two models, the procedure to show \eqref{Rel-V} is similar. Specifically for
the second model, we can use \eqref{30.02}-\eqref{2.30.0}. So, we have
\begin{eqnarray}
\label{ratioprove4}
{\mathcal{V}}^{(2)}_{j}(\pm i \lambda , a)=c^{(2)}_j \frac{M_{\kappa, \tilde{\nu}_j}(\xi a)}{M_{\kappa, \nu_j}(\xi a)} \ .
\end{eqnarray}
Taking $\lambda\rightarrow \pm i\lambda$ we obtain: 
\begin{itemize}
	\item For positive values of $j$
\begin{eqnarray}
\label{jposi}
c^{(2)}_{j}=\frac{(\pm i \lambda)}{\xi}\frac{1}{2q|j|+1}=\pm i {\rm Im}[c^{(2)}_{j}]
\end{eqnarray}
	\item For negative values of $j$
\begin{eqnarray}
\label{jnegative}
c^{(2)}_{j}=-\frac{\xi}{\pm i \lambda} (2q|j|+1)=\pm i {\rm Im}[c^{(2)}_{j}] \ .
\end{eqnarray}
\end{itemize}
Consequently, for the second model, we get
\begin{eqnarray}
\label{ratioprove5}
{\mathcal{V}}^{(2)}_{j}(\pm i \lambda , a)=c^{(2)}_j \frac{M_{\kappa, \tilde{\nu}_j}(\xi a)}
{M_{\kappa, \nu_j}(\xi a)}=\pm i{\rm Im}[{\mathcal{V}}^{(2)}_{j}(i \lambda , a)] \ .
\end{eqnarray}

Finally for the third model, we can easily show that the relation \eqref{Rel-V} holds.

\end{document}